\documentclass[lettersize,journal]{IEEEtran}

\usepackage{cite}
\usepackage{amsmath,amssymb,amsfonts}
\usepackage{algorithm}
\usepackage{array}
\usepackage[caption=false,font=normalsize,labelfont=sf,textfont=sf]{subfig}
\usepackage{graphicx,color}

\usepackage{amsthm}
\usepackage{bm}

\newtheorem{example}{Example}
\theoremstyle{remark}

\usepackage{algpseudocode}
\graphicspath{{Pictures/}}
\usepackage{stfloats}
\usepackage[table]{xcolor}
\usepackage{url}
\usepackage{hyperref}
\usepackage{textcomp}
\usepackage{booktabs}
\usepackage{verbatim}
\usepackage{tabularx}
\usepackage{multirow}
\usepackage{makecell}
\usepackage{diagbox}
\usepackage{comment}
\usepackage{float}
\usepackage{adjustbox}
\usepackage{wasysym}
\usepackage{enumitem}
\usepackage[T1]{fontenc}

\begin{document}

\title{Learning-Based Spectrum Cartography in Low Earth Orbit Satellite Networks: A Survey}

\author{Liping Tao, Xindi Tong, and Chee Wei Tan,~\IEEEmembership{Senior Member,~IEEE,}
\thanks{Liping Tao (liping.tao@163.com), Xindi Tong (to0002di@e.ntu.edu.sg), and Chee Wei Tan (cheewei.tan@ntu.edu.sg) are with the College of Computing and Data Science, Nanyang Technological University, Singapore, 639798.}
\thanks{Liping Tao and Xindi Tong contributed equally to this work.}}


\markboth{Journal of \LaTeX\ Class Files,~Vol.~14, No.~8, August~2021}%
{Shell \MakeLowercase{\textit{et al.}}: A Sample Article Using IEEEtran.cls for IEEE Journals}

\maketitle

\begin{abstract}
Low earth orbit (LEO) satellite networks are emerging as important infrastructure for global connectivity and space-based sensing, while their rapidly varying geometry, heterogeneous measurements, and dynamic propagation environments create challenges for acquiring and exploiting spatial radio information. Spectrum cartography provides a framework for constructing and updating spatial representations of radio conditions from sparse observations. This survey reviews learning-based spectrum cartography for LEO satellite networks from an integrated inference--decision perspective. We organize the literature around three representative task categories: radio map reconstruction, localization and state estimation, and map-informed resource allocation. We further examine the progression from classical interpolation, statistical estimation, and structured optimization toward learning-based approaches that exploit environmental context, relational structure, generative modeling, and adaptive information fusion. Particular attention is given to attention mechanisms for learning query- and context-dependent interactions among irregular observations, while distinguishing learned relevance from calibrated measurement reliability. Representative formulations and illustrative examples clarify these modeling principles. Finally, we synthesize cross-cutting lessons and identify open challenges in uncertainty modeling, generalization, dynamic map updating, scalable inference, realistic evaluation, and tighter coupling between radio-map inference and network decision making.
\end{abstract}


\section{INTRODUCTION}
Recent deployment of fifth-generation (5G) networks has significantly improved communication capacity, yet their dependence on terrestrial infrastructure still limits coverage in rural, remote, maritime, and disaster-affected regions. This has motivated sixth-generation (6G) systems toward seamless global connectivity and integrated sensing and communication, while the rise of AI-native platforms such as NVIDIA's Sionna \cite{chander2024sionna} further reflects the shift toward data-driven and simulation-integrated network design \cite{luo2024leo,zhang2025ai}. In this context, low earth orbit (LEO) satellite constellations have emerged as a key enabler of 6G and beyond \cite{10992253}. Compared with geostationary systems, LEO satellites operate at lower altitudes and provide reduced latency, improved link budgets, and flexible beam coverage \cite{al2022survey}. With the rapid deployment of mega-constellations such as Starlink and OneWeb \cite{mcdowell2020low}, LEO networks are evolving into highly dynamic systems with rapidly varying geometry, heterogeneous propagation conditions, intermittent visibility, and strong resource constraints, enabling applications in global broadband access, Internet of things (IoT) connectivity, positioning, and remote sensing \cite{abdelsadek2023future, hui2025review}.

\begin{figure*}[!t]
    \centering
    \includegraphics[width=0.99\linewidth]{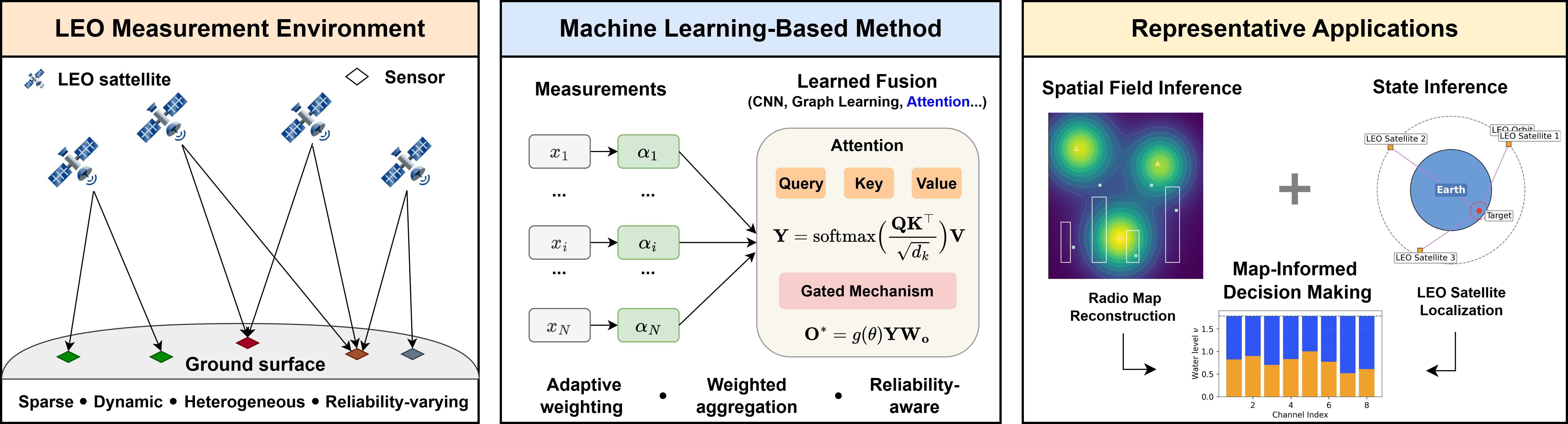}
    \caption{Unified view of learning-based spectrum cartography (SC) in low earth orbit (LEO) satellite networks. Left: sparse, dynamic, and heterogeneous LEO measurements. Middle: learning-based fusion, with attention as an important adaptive mechanism. Right: representative tasks including radio map reconstruction, LEO satellite localization, and map-informed decision making.}
    \label{fig:leo_survey_intro}
\end{figure*}

Spectrum cartography (SC) provides a natural framework for constructing spatial radio intelligence from sparse and heterogeneous observations in such environments \cite{shahaf2015information, song2026spectrum, xu2026practical}. In LEO satellite networks, SC extends beyond conventional radio-map construction and supports a broader inference-and-decision pipeline. From an application perspective, the associated tasks can be organized into three categories (cf.~Fig.~\ref{fig:leo_survey_intro}). The first is \textit{spatial field inference}, represented by radio map reconstruction, which estimates the spatial--spectral--temporal distribution of radio signals from sparse observations for spectrum sharing, interference management, and network planning \cite{liu2023uav, shrestha2023radio}. The second is \textit{state inference}, represented by LEO satellite localization, where user or satellite states are estimated from heterogeneous measurements such as range, Doppler, angle, and signal strength under rapidly varying sensing geometry \cite{wang2023doppler, shi2023revisiting, fan2024toward}. The third is \textit{map-informed decision making}, where spatial radio knowledge is further exploited for downstream tasks such as power allocation, beam switching, and beam hopping to support topology-aware and context-aware resource management in dynamic LEO networks \cite{2022leosurvey, 2024leosurveystock}.

These tasks are connected by a common need to extract useful spatial radio information from incomplete, noisy, and reliability-varying observations, although their mathematical objectives and data structures are fundamentally different \cite{sayed2005network, shibli2024data, kaplan2017understanding, langley1999dilution}. This motivates a broader \textit{learning-based SC} perspective rather than a single task-specific formulation. Depending on the sensing modality and application objective, learning-based SC can adopt different architectural biases: convolutional models are well suited to grid-structured spatial representations \cite{o2015introduction}, graph-based models capture relational or topology-dependent interactions \cite{scarselli2009graph}, while permutation-invariant set models provide a natural representation for irregular measurement collections \cite{zaheer2017deep, lee2019set}. Compared with purely model-driven or geometric interpolation approaches, learning-based methods can exploit heterogeneous contextual information and learn complex spatial dependencies that are difficult to specify analytically \cite{phillips2012practical, pesko2014radio, shibli2024data}. These complementary modeling paradigms provide the broader methodological foundation for learning-based spectrum cartography in dynamic LEO environments.

Among these paradigms, \textit{attention}-based methods are particularly relevant when observation importance varies with the query, sensing geometry, propagation conditions, or temporal context. Originating from neural machine translation \cite{bahdanau2015neural, cho2014learning} and later generalized by Transformer architectures \cite{vaswani2017attention}, attention can be interpreted as a data-adaptive weighting and aggregation mechanism, related in spirit to the Nadaraya--Watson estimator \cite{nadaraya1964estimating, watson1964smooth} but equipped with learned compatibility functions and richer contextual representations. Its permutation-compatible aggregation makes it suitable for irregular measurement sets \cite{zaheer2017deep, lee2019set}, while self-attention further enables global interactions among measurements or network entities \cite{vaswani2017attention, guo2022attention}. Recent studies have also demonstrated the potential of attention-based formulations for spatial inference from irregular radio measurements, including gridless spatial estimation and attention-based kernel regression \cite{viet2025spatial, tao2026accelerating}. Importantly, attention does not automatically provide calibrated reliability or uncertainty estimates; rather, it offers an adaptive fusion mechanism that can support reliability-aware inference when signal quality, geometry, structural priors, or other reliability-related cues are incorporated. Accordingly, this article adopts learning-based SC as its main survey scope, while treating attention as a focused methodological lens for analyzing adaptive information fusion across representative LEO applications.

\textit{Related Surveys and Research Gap.}
Existing surveys have addressed SC, localization, satellite systems, and intelligent spectrum management from complementary but largely separate perspectives. Romero et al. \cite{romero2022radio} reviewed data-driven radio map estimation, while Reddy et al. \cite{reddy2022spectrum} surveyed REM construction techniques, including compressive sensing and machine learning. For localization and satellite systems, Del Peral-Rosado et al. \cite{del2017survey} reviewed cellular localization from 1G to 5G, Prol et al. \cite{2022leosurvey} surveyed LEO-based PNT, Stock et al. \cite{2024leosurveystock} focused on opportunistic LEO-PNT, and Al-Hraishawi et al. \cite{al2022survey} provided a broad survey of NGSO satellite communications. Abdel Khalek et al. \cite{khalek2024advances} further reviewed ML-driven cognitive radio for spectrum sensing, resource management, and adaptive decision making.
As summarized in Table~\ref{tab:related_surveys}, these works provide strong coverage of individual areas, but do not jointly organize radio map reconstruction, LEO state inference, and downstream network decision making within a unified spectrum-cartography framework. To the best of our knowledge, a systematic survey connecting these tasks from a learning-based LEO perspective remains missing. This survey addresses this gap and further highlights attention and adaptive fusion as an important methodological thread for handling irregular, geometry-dependent, and heterogeneous measurements.

\begin{table*}[!t]
\centering
\caption{Comparison of representative surveys related to spectrum cartography, localization, satellite systems, and learning-enabled spectrum management.}
\label{tab:related_surveys}
\renewcommand{\arraystretch}{1.15}
\setlength{\tabcolsep}{5pt}
\begin{adjustbox}{width=\textwidth,center}
\begin{tabular}{
    p{1.4cm}
    p{3,9cm}
    c c c c c c c
}
\hline
\textbf{Reference}
&
\textbf{Main Scope}
&
\textbf{LEO/ NGSO}
&
\textbf{Radio Map/ SC}
&
\textbf{Localization/ PNT}
&
\textbf{Learning Methods}
&
\textbf{Map-Informed Resource Mgmt.}
&
\textbf{Attention/ Adaptive Fusion}
&
\textbf{Cross-Task View}
\\
\hline

\cite{romero2022radio}
&
Radio map estimation and SC
&
$\pmb-$
&
$\pmb{\checkmark}$
&
$\pmb\circ$
&
$\pmb{\checkmark}$
&
$\pmb\circ$
&
$\pmb-$
&
$\pmb-$
\\

\cite{reddy2022spectrum}
&
SC and REM construction
&
$\pmb-$
&
$\pmb{\checkmark}$
&
$\pmb\circ$
&
$\pmb{\checkmark}$
&
$\pmb\circ$
&
$\pmb-$
&
$\pmb-$
\\

\cite{del2017survey}
&
Cellular localization
&
$\pmb-$
&
$\pmb-$
&
$\pmb{\checkmark}$
&
$\pmb\circ$
&
$\pmb\circ$
&
$\pmb-$
&
$\pmb-$
\\

\cite{2022leosurvey}
&
LEO-PNT systems
&
$\pmb{\checkmark}$
&
$\pmb-$
&
$\pmb{\checkmark}$
&
$\pmb\circ$
&
$\pmb\circ$
&
$\pmb-$
&
$\pmb-$
\\

\cite{2024leosurveystock}
&
Opportunistic LEO-PNT
&
$\pmb{\checkmark}$
&
$\pmb-$
&
$\pmb{\checkmark}$
&
$\pmb\circ$
&
$\pmb-$
&
$\pmb-$
&
$\pmb-$
\\

\cite{al2022survey}
&
NGSO satellite communications
&
$\pmb{\checkmark}$
&
$\pmb-$
&
$\pmb\circ$
&
$\pmb\circ$
&
$\pmb{\checkmark}$
&
$\pmb-$
&
$\pmb-$
\\

\cite{khalek2024advances}
&
ML-driven cognitive radio
&
$\pmb-$
&
$\pmb\circ$
&
$\pmb\circ$
&
$\pmb{\checkmark}$
&
$\pmb{\checkmark}$
&
$\pmb\circ$
&
$\pmb\circ$
\\

\textbf{This survey}
&
\textbf{Learning-based SC for LEO networks}
&
$\pmb{\checkmark}$
&
$\pmb{\checkmark}$
&
$\pmb{\checkmark}$
&
$\pmb{\checkmark}$
&
$\pmb{\checkmark}$
&
$\pmb{\checkmark}$
&
$\pmb{\checkmark}$
\\
\hline

\multicolumn{9}{l}{
\footnotesize
$\pmb{\checkmark}$: substantial coverage;
$\pmb\circ$: partial or related coverage;
$\pmb-$: not a primary focus.
}

\end{tabular}
\end{adjustbox}

\end{table*}

\textit{Scope and Literature Selection.}
This survey focuses on learning-based SC for LEO satellite networks, covering spatial field inference (e.g., radio map reconstruction), state inference (e.g., LEO-assisted localization), and map- or context-informed network decision making (e.g., spectrum, power, and beam resource management). Classical estimation, interpolation, and optimization methods are included when they provide methodological foundations or meaningful baselines, while attention is treated as an important adaptive-fusion mechanism within the broader learning-based framework. Relevant studies published up to 2026 were identified through structured searches of IEEE Xplore, Scopus, Web of Science, and Google Scholar using combinations of LEO/satellite-related terms, spectrum-cartography and spatial-inference terms, learning-related terms, and application-specific keywords such as localization, positioning, resource allocation, beam hopping, spectrum management, and power allocation; backward and forward citation tracing was also used. Studies were included if they addressed learning-based radio map or spectrum-cartography construction, learning-enabled LEO localization or measurement fusion, spatial-radio-information-driven network decision making, or methodological advances relevant to adaptive fusion of sparse and heterogeneous measurements.

\textit{Contribution.}
This article provides a systematic survey of learning-based spectrum cartography for LEO satellite networks, covering spatial field inference, state inference, and map- or context-informed network decision making. The main contributions are summarized as follows:
\begin{itemize}
    \item We establish an LEO-oriented spectrum-cartography framework that organizes representative tasks into spatial field inference, state inference, and downstream map-informed decision making, thereby connecting sensing, spatial radio intelligence, and network control within a common inference--decision perspective.

    \item We systematically review learning-based methods for radio map reconstruction, LEO localization and positioning, and map-informed resource management, and compare their measurement representations, learning paradigms, use of side information, reliability handling, and evaluation settings.

    \item We provide a focused synthesis of attention and adaptive fusion as an important methodological thread for irregular and heterogeneous measurements, clarifying their relationship with classical weighting methods and their roles and limitations in reliability-aware inference.

    \item We distill cross-cutting lessons and identify open research challenges in uncertainty modeling, physics--learning integration, generalization, scalability, onboard and distributed inference, and real-world evaluation. The source code is publicly available.\footnote{Source code is available at \url{https://github.com/convexsoft/LearnSCLEO}.}
\end{itemize}

\begin{figure*}[!t]
    \centering
    \includegraphics[width=0.99\linewidth]{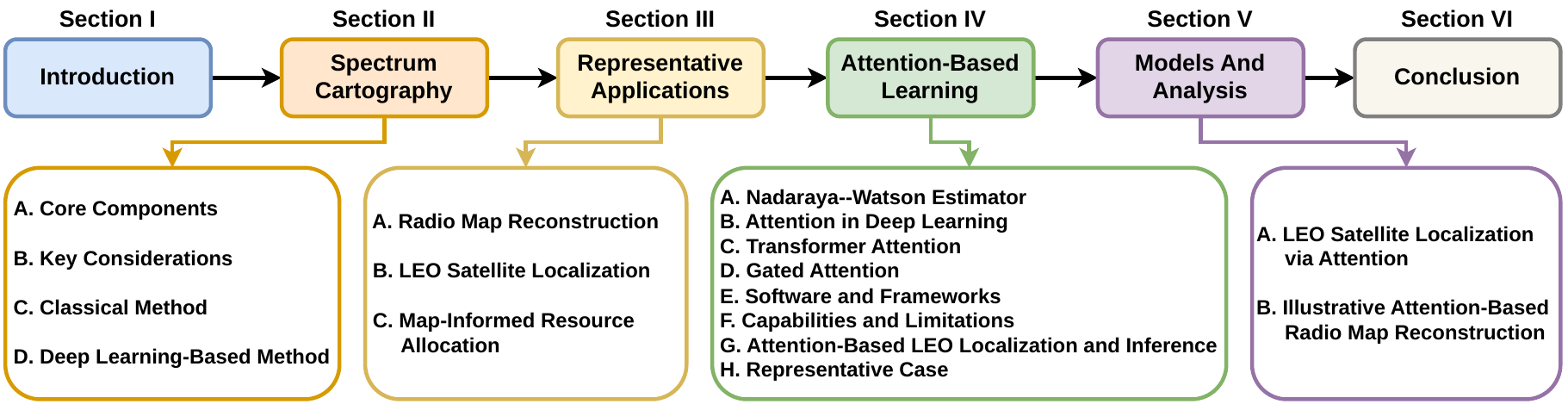}
    \caption{Organization of the paper, progressing from spectrum cartography fundamentals and representative applications to attention-based models and analysis.}
    \label{fig:sc_leo_organization}
\end{figure*}

\textit{Organization.}
As shown in Fig.~\ref{fig:sc_leo_organization}, Section~\ref{sec:sc} introduces the foundations of spectrum cartography, including its core components/lifecycle, key considerations, and classical and learning-based methods. Section~\ref{sec:sc-background-rem-leo} reviews three representative LEO applications: radio map reconstruction, localization and state estimation, and map-informed resource allocation. Section~\ref{sec:attention-leo-sc} reviews attention mechanisms for adaptive information fusion and discusses their capabilities, limitations, and roles in LEO spectrum cartography tasks. Section~\ref{sec:attention-leo-rem-model-method-simu} presents illustrative attention-based formulations for LEO localization and radio map reconstruction to clarify the transition from fixed weighting to learned, context-dependent aggregation. Finally, Section~\ref{sec:conclusion} concludes the paper and summarizes key future research directions.

\section{SPECTRUM CARTOGRAPHY}
\label{sec:sc}
Spectrum cartography aims to infer spatially continuous radio quantities, such as received signal strength (RSS), pathloss, channel gain, interference power, or power spectral density (PSD), from sparse and geo-referenced observations \cite{romero2022radio,romero2017learning}. The resulting radio maps characterize spectrum usage and propagation conditions across space, frequency, and time, and support applications such as spectrum sharing, interference management, and network planning \cite{Dynamic2015chen,Shrestha2022DeepSC,reddy2022spectrum}. SC has evolved from classical spatial interpolation and kernel-based methods \cite{oliver1990kriging} toward structured inverse formulations and, more recently, learning-based approaches that exploit heterogeneous measurements and environmental context \cite{timilsina2025domain,jaensch2026radiomapnoisy}. This section introduces the SC lifecycle, its main methodological principles, and the task taxonomy adopted throughout this survey.

\subsection{Spectrum Cartography Lifecycle}
\begin{figure*}[t]
    \centering
    \includegraphics[width=0.88\linewidth]{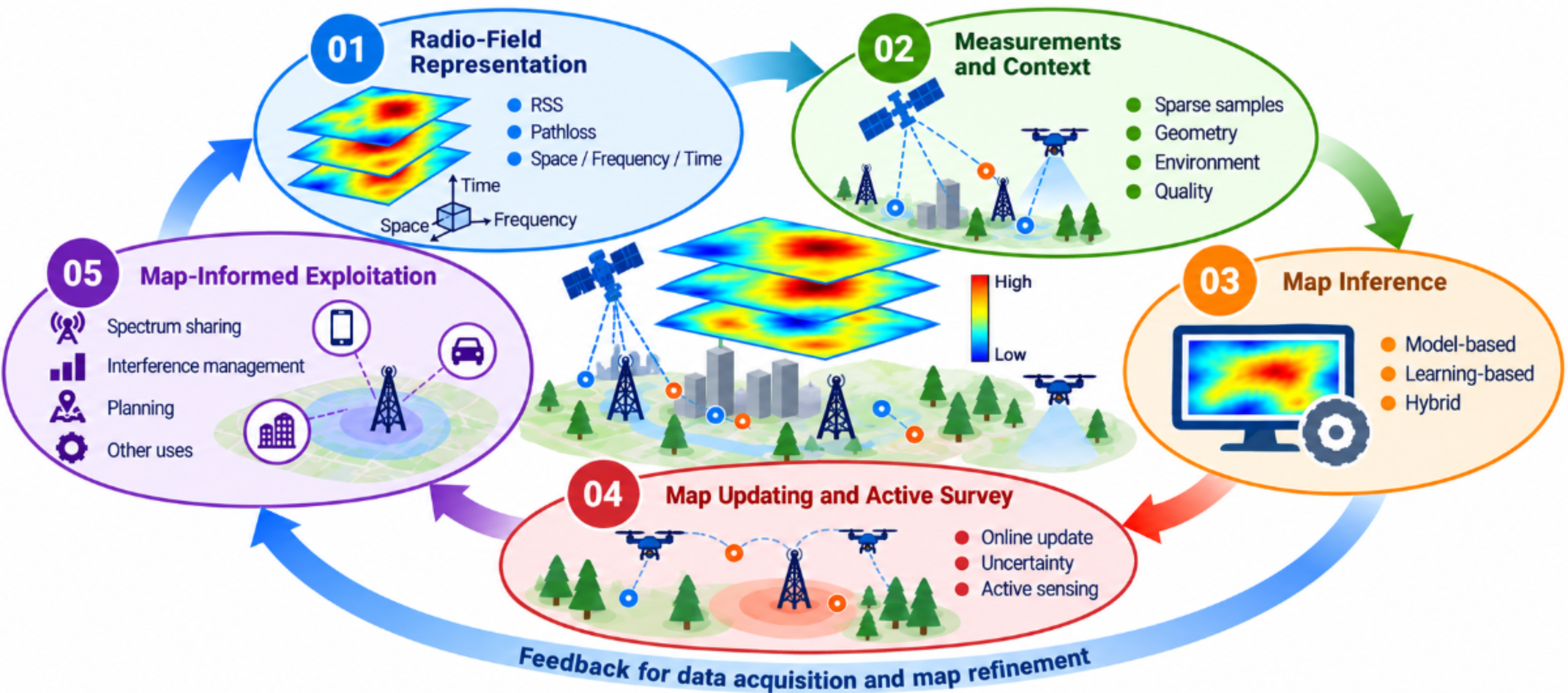}
    \caption{Core components and lifecycle of spectrum cartography. Sparse measurements and contextual information are used to infer and update radio-field representations through model-based, learning-based, or hybrid methods. The resulting maps support downstream network decisions, while feedback from current estimates guides subsequent data acquisition and refinement.}
    \label{fig:sc_lifecycle}
\end{figure*}

As illustrated in Fig.~\ref{fig:sc_lifecycle} and Table ~\ref{tab:sc_components}, the SC lifecycle can be described through five interrelated components: radio-field representation, measurements and context, map inference, map updating and active surveying, and map-informed exploitation. They describe how a target radio quantity is defined, observed, reconstructed, refined, and ultimately used for network analysis and decision making \cite{romero2022radio,reddy2022spectrum}.

\subsubsection{Radio-Field Representation}
Radio-field representation specifies the target quantity and modeling domain. Typical targets include RSS, pathloss, channel gain, interference power, and PSD \cite{romero2022radio}. Depending on the application, the field may be represented over space alone or extended to space--frequency and space--time domains \cite{Dynamic2015chen}. This choice determines the dimensionality and structure of the subsequent inference problem.

\subsubsection{Measurements and Context}
SC relies on sparse, heterogeneous observations and contextual information for radio-field inference. Classical formulations primarily use geo-referenced signal measurements \cite{Dynamic2015chen,romero2015stochastic,ids2017spectrum}, while recent learning-based methods additionally exploit transmitter information, environmental geometry, building maps, aerial imagery, and other side information \cite{levie2021radiounet,jaensch2026radiomapaerial}. Location-free and geometry-aware formulations further relax the dependence on accurate sensor coordinates or incorporate directional propagation effects \cite{teganya2019location,maleki2016cooperative}. These developments highlight that measurement quality, geometry, and contextual information can be as important as the measured signal values themselves.

\subsubsection{Map Inference}
Map inference transforms incomplete observations into a continuous estimate of the radio field. Representative approaches include kernel regression \cite{scholkopf2002learning,tao2026accelerating}, semiparametric modeling \cite{romero2015stochastic}, adaptive basis expansions \cite{ids2017spectrum}, low-rank or tensor-based reconstruction \cite{Shrestha2022DeepSC}, and learning-based predictors \cite{levie2021radiounet,jaensch2026radiomapnoisy}. Although these methods differ in assumptions and model structure, they share the objective of recovering unobserved spatial radio information from partial evidence.

\subsubsection{Map Updating and Active Surveying}
Spectrum cartography extends beyond one-shot reconstruction by maintaining and refining radio maps as new observations become available \cite{romero2022radio}. Online updating tracks temporal variations in the radio environment, while active surveying adaptively selects new sensing locations based on the current map estimate, uncertainty, or expected information gain \cite{shrestha2023spectrum}. This creates a feedback loop between inference and data acquisition, enabling progressive map refinement under limited sensing resources.

\subsubsection{Map-Informed Exploitation}
The inferred radio map can subsequently support network analysis and decision making. Typical applications include spectrum sharing, interference management, coverage assessment, network planning, and resource allocation \cite{romero2022radio,reddy2022spectrum}. Recent approaches increasingly integrate radio-map prediction with optimization and control, allowing learned spatial radio information to directly inform system-level decisions \cite{jaensch2026radiomapnoisy}. In this sense, SC provides a bridge from sparse measurements to actionable radio-environment knowledge.

\subsection{Key Considerations for Radio-Map Inference}
We summarize radio-map inference through three closely related considerations: coherence, coverage, and structure, inspired by the notion of information cartography~\cite{shahaf2015information}. Together, they characterize how spatial dependence, measurement distribution, and propagation-aware context affect the recoverability of a radio field from sparse observations \cite{romero2022radio}.

\subsubsection{Coherence}
Coherence refers to the spatial dependence of radio quantities, whereby nearby locations often exhibit correlated propagation characteristics due to large-scale effects such as pathloss and shadowing. This property underlies classical interpolation and kernel-based methods and enables information from observed locations to generalize to nearby unobserved regions \cite{oliver1990kriging,romero2015stochastic}. However, spatial proximity alone does not guarantee similarity when strong blockage or environmental transitions are present. For example, if two nearby measurements have similar RSS values, the value at an intermediate location is typically expected to vary smoothly unless a strong obstruction intervenes.

\subsubsection{Coverage}
Coverage describes how well the available measurements span the region of interest. Sparse or highly uneven sampling leaves poorly observed regions in which prediction increasingly relies on extrapolation rather than interpolation, typically increasing reconstruction uncertainty. Measurement placement therefore directly affects radio-map quality and motivates adaptive surveying and active sensing strategies \cite{shrestha2023spectrum}. For instance, measurements concentrated along a narrow trajectory may support accurate interpolation nearby but provide unreliable predictions in distant unsampled regions.

\subsubsection{Structure}
Structure captures propagation-dependent regularities that cannot be explained by spatial smoothness alone, including blockage, terrain, transmitter geometry, antenna directionality, and environmental context. Modern learning-based SC increasingly incorporates such information through environmental maps, auxiliary features, graph relations, or other structured representations \cite{levie2021radiounet,maleki2016cooperative,jaensch2026radiomapaerial}. For example, two geographically close locations separated by a building may exhibit substantially different received powers despite their spatial proximity.

\subsection{Classical Methods}
Spectrum cartography aims to reconstruct spatial, spectral, or joint spatio-spectral radio maps from geographically distributed yet typically sparse measurements. A generic observation model can be written as:
\begin{equation}
\mathbf{Y}=\mathcal{P}_{\Omega}(\mathbf{X})+\mathbf{N},
\end{equation}
where $\mathbf{X}$ is the complete radio map, $\Omega$ denotes the set of sampled locations, $\mathcal{P}_{\Omega}(\cdot)$ is the sampling operator, and $\mathbf{N}$ denotes measurement noise. Classical methods address this ill-posed inverse problem by imposing handcrafted priors such as smoothness, sparsity, or low rank \cite{Shrestha2022DeepSC,Dynamic2015chen}. This line of research traces back to early interference cartography and radio map construction, where spatial maps of interference or received power were built from geo-localized measurements for cognitive radio applications \cite{ben2008Informed}.

\textit{Kriging.}
A representative classical approach is Kriging, which models the radio map as a random field and yields a linear minimum mean-square error estimator \cite{ben2008ICH,ben2008Informed,romero2022radio}:
\begin{equation}
\hat{p}(\mathbf{s}_0)
=
\mu(\mathbf{s}_0)
+\mathbf{k}(\mathbf{s}_0)^{\top}
\big(\mathbf{K}+\sigma_n^2\mathbf{I}\big)^{-1}
(\mathbf{y}-\boldsymbol{\mu}),
\end{equation}
where $\mu(\cdot)$ is the mean function, $\mathbf{K}$ is the covariance matrix over sampled locations, and $\mathbf{k}(\mathbf{s}_0)$ contains cross-covariances with the target location $\mathbf{s}_0$. Kriging is attractive for its statistical interpretability and uncertainty quantification, but its performance depends strongly on the assumed covariance model \cite{Shrestha2022DeepSC}.

\textit{Kernel Methods.}
Another important family includes kernel and basis-expansion methods such as thin-plate splines, reproducing kernel Hilbert space regression \cite{tao2026accelerating}, and radial basis function interpolation \cite{romero2022radio,ids2017spectrum}, where the map is expressed as:
\begin{equation}
\hat{p}(\mathbf{s})=\sum_{m=1}^{M}\alpha_m\,\kappa(\mathbf{s},\boldsymbol{\mu}_m).
\end{equation}
Here, $\kappa(\cdot,\cdot)$ is a kernel, $\boldsymbol{\mu}_m$ are kernel centers, and $\alpha_m$ are coefficients estimated from the measurements. These methods adapt naturally to irregular sampling, although their performance still depends on the chosen kernel family.

\textit{Structured Recovery.}
A third class formulates spectrum cartography as structured recovery, including compressive sensing, dictionary learning \cite{kim2013cognitive}, matrix completion \cite{schaufele2019tensor}, and tensor completion \cite{Shrestha2022DeepSC,Dynamic2015chen}. For spatio-spectral maps, one may write:
\begin{equation}
\mathbf{X}\approx\sum_{r=1}^{R}\mathbf{S}_r\circ\mathbf{c}_r,
\label{eq:sp-spe}
\end{equation}
where $\mathbf{S}_r$ and $\mathbf{c}_r$ denote the spatial loss field and spectral signature of emitter $r$, respectively, motivating matrix/tensor recovery that exploits spatial and spectral correlations.

Overall, classical methods remain valuable for their interpretability, statistical foundations, and effectiveness with limited data. However, their reliance on handcrafted priors limits their flexibility in complex propagation environments, motivating learning-based spectrum cartography.

\subsection{Deep Learning-Based Methods}
Deep learning-based spectrum cartography replaces handcrafted priors with representations learned directly from data. The central idea is that radio maps lie on a low-dimensional manifold that can be learned from training samples and used to infer missing values from sparse measurements \cite{romero2022radio}. This is particularly effective in complex propagation environments, where assumptions such as smoothness or low rank may be overly restrictive \cite{Shrestha2022DeepSC}.

A representative starting point is the deep completion paradigm, where a network reconstructs a complete radio map from a masked input. Given a complete training map $\mathbf{Q}_n$ and a sampling mask $\mathbf{M}_n$, training can be formulated as:
\begin{equation}
\hat{\theta}
=
\arg\min_{\theta}
\frac{1}{N}\sum_{n=1}^{N}
\left\|
f_{\theta}\!\left(\mathbf{M}_n \odot \mathbf{Q}_n\right)-\mathbf{Q}_n
\right\|_F^2,
\end{equation}
where $f_{\theta}(\cdot)$ is the completion network. In encoder--decoder form, $f_{\theta}(\mathbf{Q}) = g_{\theta_d}\!\left(p_{\theta_e}(\mathbf{Q})\right)$, the latent representation provides a compact description of the radio-map manifold and supports interpolation from sparse observations \cite{Shrestha2022DeepSC,romero2022radio}.

\begin{table*}[t]
\centering
\caption{Core components and lifecycle of spectrum cartography.}
\label{tab:sc_components}

\setlength{\tabcolsep}{4pt}
\setlength{\aboverulesep}{0pt}
\setlength{\belowrulesep}{0pt}
\setlength{\extrarowheight}{1.5pt}
\renewcommand{\arraystretch}{1.0}

\begin{tabular}{|>{\raggedright\arraybackslash}m{3.15cm}|
                >{\raggedright\arraybackslash}m{9.55cm}|
                >{\raggedright\arraybackslash}m{4.45cm}|}
\toprule
\rowcolor{gray!15}
\textbf{Component} & \textbf{Key Roles and Elements} & \textbf{Representative Work} \\
\midrule

\textbf{Radio-field representation}
&
\begin{itemize}[leftmargin=*, itemsep=2pt, topsep=2pt, parsep=0pt, partopsep=0pt]
    \item \textbf{Target:} RSS, pathloss, channel gain, interference, or PSD
    \vspace{-1.5mm}
    \item \textbf{Domain:} space, frequency, and time
    \vspace{-1.5mm}
    \item \textbf{Role:} defines the structure of the target radio field
\end{itemize}
& \cite{romero2022radio,Dynamic2015chen} \\
\midrule

\textbf{Measurements \& context}
&
\begin{itemize}[leftmargin=*, itemsep=2pt, topsep=2pt, parsep=0pt, partopsep=0pt]
    \item \textbf{Measurements:} sparse and heterogeneous signal observations
    \vspace{-1.5mm}
    \item \textbf{Context:} geometry, transmitter information, and environmental data
    \vspace{-1.5mm}
    \item \textbf{Quality:} location uncertainty, noise, and incomplete observations
\end{itemize}
& \cite{teganya2019location,levie2021radiounet,jaensch2026radiomapaerial,jaensch2026radiomapnoisy} \\
\midrule

\textbf{Map inference}
&
\begin{itemize}[leftmargin=*, itemsep=2pt, topsep=2pt, parsep=0pt, partopsep=0pt]
    \item \textbf{Tasks:} reconstruction, completion, and prediction
    \vspace{-1.5mm}
    \item \textbf{Methods:} model-based, learning-based, and hybrid approaches
    \vspace{-1.5mm}
    \item \textbf{Goal:} recover unobserved radio information from partial evidence
\end{itemize}
& \cite{Dynamic2015chen,romero2015stochastic,ids2017spectrum,Shrestha2022DeepSC} \\
\midrule

\textbf{Map updating \& active surveying}
&
\begin{itemize}[leftmargin=*, itemsep=2pt, topsep=2pt, parsep=0pt, partopsep=0pt]
    \item \textbf{Updating:} refine maps as new observations become available
    \vspace{-1.5mm}
    \item \textbf{Surveying:} adaptively select new sensing locations
    \vspace{-1.5mm}
    \item \textbf{Feedback:} use uncertainty or information gain to guide acquisition
\end{itemize}
& \cite{romero2022radio,shrestha2023spectrum} \\
\midrule

\textbf{Map-informed exploitation}
&
\begin{itemize}[leftmargin=*, itemsep=2pt, topsep=2pt, parsep=0pt, partopsep=0pt]
    \item \textbf{Applications:} spectrum sharing and interference management
    \vspace{-1.5mm}
    \item \textbf{Planning:} coverage analysis, network design, and resource allocation
    \vspace{-1.5mm}
    \item \textbf{Integration:} optimization and control using inferred radio maps
\end{itemize}
& \cite{romero2022radio,reddy2022spectrum,jaensch2026radiomapnoisy} \\
\bottomrule
\end{tabular}
\end{table*}

To reduce the difficulty of learning aggregated radio maps, subsequent work introduces emitter-wise decomposition as in (\ref{eq:sp-spe}). By learning individual components rather than the aggregated tensor directly, this approach improves generalization while preserving physical structure \cite{Dynamic2015chen}.

Among practical architectures, convolutional neural networks (CNNs) have become the dominant backbone for radio map prediction. They treat environmental information, transmitter locations, and sparse measurements as image-like inputs and learn:
\begin{equation}
\hat{\mathbf{X}} =
f_{\theta}\!\left(
\mathbf{I}_{\mathrm{env}},
\mathbf{I}_{\mathrm{Tx}},
\mathbf{I}_{\mathrm{meas}}
\right).
\end{equation}

A prominent example is RadioUNet, which uses a U-Net-style encoder--decoder to infer path-loss maps from city maps and transmitter positions \cite{levie2021radiounet}. Subsequent studies further explore benchmark datasets, aerial-image-based prediction, and robustness to imperfect side information \cite{jaensch2026radiomapaerial,jaensch2026radiomapnoisy}. Meanwhile, lightweight data-driven schemes such as kernel-density-based learning have also been investigated, highlighting the tradeoff between model expressiveness and computational complexity \cite{xu2021efficient}.

In summary, learning-based spectrum cartography has evolved from neural completion to structured decomposition and CNN-based scene-aware prediction, and is now moving toward more expressive models that capture global spatio-temporal dependencies and heterogeneous observations \cite{yapar2025sampling}. In this context, attention- and Transformer-based architectures extend the representational scope of CNNs and have emerged as a key direction for next-generation spectrum cartography \cite{Shrestha2022DeepSC,romero2022radio,Dynamic2015chen}.

\section{REPRESENTATIVE APPLICATIONS}
\label{sec:sc-background-rem-leo}
\subsection{Radio Map Reconstruction}
\begin{figure}
    \centering
    \includegraphics[width=0.94\linewidth]{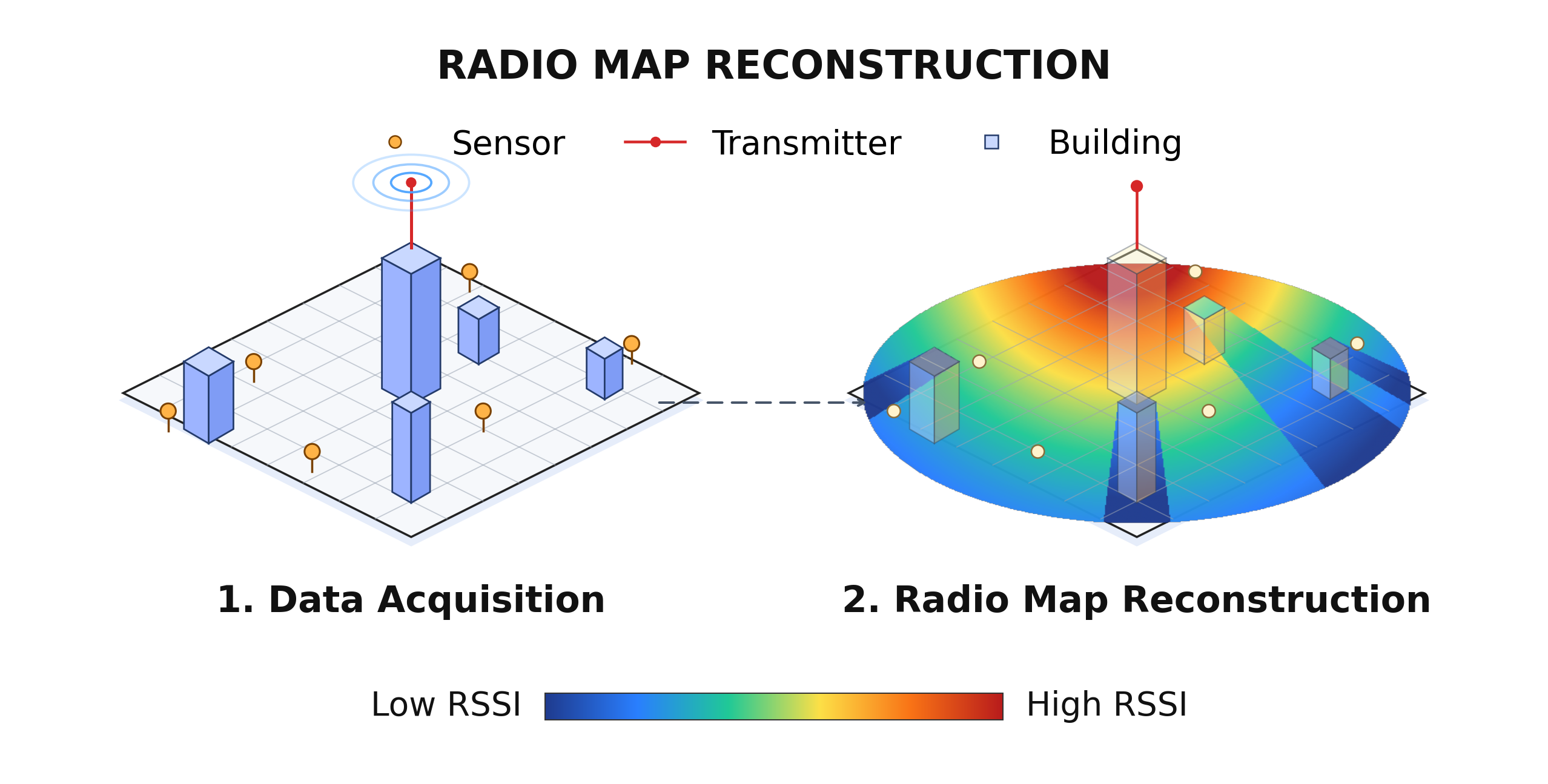}
    \caption{Illustration of map-informed resource allocation in LEO satellite networks. Spatial radio information inferred from sparse observations is organized into a radio map and subsequently used to support downstream decisions such as power and spectrum allocation, beam selection, and spatiotemporal beam scheduling.}
    \label{fig:radio_map}
\end{figure}

Radio map reconstruction estimates spatially varying radio quantities, such as RSS, reference signal received power (RSRP), channel gain, interference power, or PSD, from sparse observations over space, frequency, and time \cite{romero2022radio,reddy2022spectrum}. As illustrated in Fig.~\ref{fig:radio_map}, the objective is to recover unobserved radio conditions from incomplete measurements while accounting for propagation effects such as blockage, shadowing, and spatial heterogeneity. Recent learning-based approaches further incorporate environmental context, transmitter information, and heterogeneous sensing features, moving beyond purely distance-based interpolation toward data-driven spatial inference \cite{bi2019engineering,li2025secure,chen2025gaussian}.

\subsubsection{Problem Formulation}
Let $\mathcal{D}=\left\{\left(\mathbf{x}_i,\mathbf{m}_i,\mathbf{c}_i\right)\right\}_{i=1}^{N_s}$ denote a set of sparse observations, where $\mathbf{x}_i$ is the measurement location, $\mathbf{m}_i$ contains the observed radio quantities, and $\mathbf{c}_i$ denotes optional contextual information such as frequency, time, transmitter descriptors, or environmental features. Radio map reconstruction seeks to estimate the radio field at an arbitrary query location $\mathbf{x}_q$ as:
\begin{equation}
\hat{F}(\mathbf{x}_q)
=
\mathcal{M}_{\theta}
\left(
\mathbf{x}_q;\mathcal{D}
\right),
\label{eq:radio_map_general}
\end{equation}
where $\mathcal{M}_{\theta}$ denotes a model-based, learning-based, or hybrid estimator. Depending on the application, $F$ may represent a spatial field or a joint space--frequency--time field. The central challenge is to infer unobserved radio conditions from sparse, noisy, and spatially heterogeneous evidence.

\subsubsection{Fundamental Challenges}
Radio map reconstruction faces three closely related challenges: sparse and uneven sensing, propagation heterogeneity, and environmental dynamics. First, measurements are typically sparse, nonuniformly distributed, and collected from heterogeneous sensing sources with different noise characteristics \cite{dare2023radio}. Limited or clustered sampling leaves large regions weakly observed and makes reconstruction inherently ill-posed, motivating the use of spatial structure, contextual information, and appropriate inductive priors \cite{bi2019engineering,wang2024sparse}.
Second, radio fields are governed by propagation physics and environmental geometry rather than spatial proximity alone. Blockage, shadowing, transmitter geometry, and antenna characteristics can induce anisotropic and highly context-dependent correlations, such that nearby locations may exhibit substantially different radio conditions \cite{liu2023uav}. Consequently, purely distance-based interpolation or fixed smoothing rules may oversmooth sharp propagation transitions or fail to capture geometry-dependent variations \cite{phillips2012practical}.

Finally, reconstruction must cope with time-varying radio environments and model uncertainty. In LEO satellite networks, rapid satellite motion, changing visibility and beam footprints, and evolving propagation conditions can shift both the target radio field and the measurement distribution over time. Reconstruction models therefore need to generalize across changing sensing geometries and propagation regimes while accounting for spatially varying prediction uncertainty \cite{shrestha2023radio,shrestha2023spectrum}. These challenges motivate learning-based approaches that combine flexible representations with contextual, structural, and uncertainty-related information.

\subsubsection{Classical Reconstruction Methods}
Radio map reconstruction has been addressed through a broad range of model-based, statistical, and learning-based approaches. Classical methods remain important methodological foundations and useful baselines, particularly when training data are limited or reliable propagation/statistical priors are available \cite{romero2022radio,reddy2022spectrum,pesko2014radio}.

\textit{Classical Foundations.}
Direct interpolation methods estimate unobserved radio quantities from neighboring measurements using spatial proximity or statistical correlation. Representative approaches include inverse distance weighting (IDW) \cite{shepard1968two}, nearest-neighbor and spline interpolation \cite{fix1985discriminatory,de1978practical}, natural-neighbor interpolation \cite{sibson1981brief}, and Kriging \cite{oliver1990kriging}. These methods are simple and data-efficient but may become inaccurate under nonuniform sampling or abrupt propagation changes. Model-based approaches instead incorporate propagation laws or environmental knowledge through parameterized radio models, providing physical interpretability but remaining sensitive to model mismatch \cite{pesko2014radio}. Structured and probabilistic approaches further exploit sparsity, low-rank structure, or statistical priors through compressive sensing, matrix/tensor completion, Gaussian-process-type inference, and Bayesian estimation \cite{Dynamic2015chen,Shrestha2022DeepSC,wang2024sparse}. Such methods can improve reconstruction under limited observations and, in probabilistic formulations, provide explicit uncertainty estimates, although their performance still depends on the validity of the assumed structure or prior.

\subsubsection{Learning-Based Reconstruction Methods}
Recent learning-based approaches move beyond fixed interpolation rules and handcrafted spatial priors by learning propagation structures, environmental dependencies, or measurement interactions directly from data. Existing methods differ mainly in how they represent radio fields and observations, what inductive biases they impose, and whether they provide deterministic or probabilistic predictions. For clarity, we organize representative approaches into five families: grid-based completion and convolutional models, geometry- and context-aware models, graph- and structured learning models, generative and probabilistic models, and attention- and Transformer-based models.

\textit{Grid-Based Completion and Convolutional Models.}
Early deep-learning approaches commonly discretize the region of interest onto a regular grid and formulate radio map reconstruction as image or tensor completion. Teganya et al. \cite{teganya2022deep} developed a fully convolutional completion autoencoder that learns recurring spatial propagation structures from previously observed radio maps, thereby reducing the number of measurements required during reconstruction. RadioUNet further adopts a UNet architecture to predict pathloss maps from urban geometry and transmitter locations, providing fast approximations to computationally intensive propagation simulations \cite{levie2021radiounet}. More recent work improves data efficiency through self-supervised pretraining, where masked environmental or radio-map information is reconstructed before task-specific fine-tuning \cite{ma2025selfsupervised}. These approaches benefit from the strong spatial inductive bias and computational efficiency of convolution, but their reliance on rasterized inputs and fixed grids may limit spatial resolution and make irregular measurement sets less natural to represent.

\textit{Geometry- and Context-Aware Models.}
A complementary direction explicitly incorporates propagation-relevant context, including building geometry, transmitter descriptors, aerial imagery, material information, and other environmental features. Geography-aware approaches exploit environmental structure to model blockage and propagation effects that cannot be captured by spatial proximity alone \cite{liu2023uav}. CNN-based methods further combine such context with sparse signal observations, while recent studies investigate robustness when environmental information is incomplete or inaccurate. In particular, Jaensch et al. treat environment information as an uncertain prior and sparse radio measurements as spatial anchors, and show that training with physically motivated perturbations improves robustness to errors in geometry, material properties, and transmitter locations \cite{jaensch2026radiomapnoisy}. These methods demonstrate the value of side information for propagation-aware inference, but their performance may depend strongly on the availability, accuracy, and transferability of environmental context.

\textit{Graph- and Structured Learning Models.}
When measurements or network entities are naturally relational rather than grid-aligned, graph-based representations provide an alternative to rasterization. Shibli and Zanouda encode network coverage regions as a graph and employ graph neural networks to fuse physical cell information with sparse geo-located RSRP/RSRQ measurements, allowing spatial dependencies to be learned over neighboring regions \cite{shibli2024data}. Structured hybrid approaches instead combine learned representations with explicit radio-map decompositions. For example, Shrestha et al. model individual emitters with learned neural models and reconstruct the aggregate spatio-spectral map through a structured completion framework, reducing the effective state space and alleviating training and out-of-distribution generalization difficulties \cite{Shrestha2022DeepSC}. Such approaches impose stronger relational or physical structure than generic convolutional models, but their performance depends on the suitability of the graph construction, decomposition, or structural assumptions.

\textit{Generative and Probabilistic Models.}
Generative models capture the distribution or latent structure of radio fields, enabling flexible reconstruction under sparse and incomplete observations. GAN-based methods include ACT-GAN, which incorporates contextual and attention modules for improved reconstruction consistency \cite{chen2025actgan}, and 3D-DCRGAN, which employs unsupervised adversarial learning and dilated convolutions for three-dimensional radio map reconstruction \cite{hu2023gan3d}. Probabilistic approaches, such as RadioVAE \cite{yang2025radiovae} and Gaussian-process-based methods \cite{zhen2022radio}, further characterize prediction uncertainty. More recently, diffusion-based methods have been explored for sparse radio map reconstruction, including TSRDiff, which combines conditional diffusion with environmental information and attention-based feature extraction \cite{ye2025adiffusion}, and communication-aware diffusion reconstruction for dynamic aerial networks \cite{liu2026diffusion}. Despite their flexibility in complex propagation environments, generative approaches generally involve higher computational costs, particularly diffusion models requiring iterative denoising.

\textit{Attention- and Transformer-Based Models.}
Attention-based models provide a flexible alternative for reconstructing radio fields from irregular measurement collections. Rather than first projecting observations onto a fixed grid, attention can directly learn query-dependent interactions between measurement locations and prediction points. A representative example is STORM, a spatial Transformer for radio map estimation that operates in a gridless manner, preserves full spatial resolution, and reduces the parameter count relative to conventional convolutional estimators \cite{viet2025spatial}. Its attention-based formulation further supports translation- and rotation-equivariant estimation and can be extended to active sensing. Attention modules have also been embedded within convolutional and generative architectures to enhance contextual feature extraction, whereas Transformer-based methods such as STORM use attention as the primary inference mechanism. Although attention is particularly useful when measurement relevance varies with query location or geometry, its weights indicate learned relevance rather than calibrated measurement reliability. Further discussion is provided in Section~\ref{sec:attention-leo-sc}.

\begin{table*}[!t]
\centering
\caption{Representative learning-based methods for radio map reconstruction.}
\label{tab:learning_radiomap}

\scriptsize
\renewcommand{\arraystretch}{1.05}
\setlength{\tabcolsep}{3.2pt}
\setlength{\aboverulesep}{0pt}
\setlength{\belowrulesep}{0pt}
\setlength{\extrarowheight}{1.5pt}

\begin{tabular}{
|>{\raggedright\arraybackslash}p{1.2cm}|
>{\raggedright\arraybackslash}p{1.7cm}|
>{\raggedright\arraybackslash}p{3.0cm}|
>{\raggedright\arraybackslash}p{2.05cm}|
>{\raggedright\arraybackslash}p{1.75cm}|
>{\raggedright\arraybackslash}p{1.35cm}|
>{\raggedright\arraybackslash}p{4.95cm}|}
\toprule
\rowcolor{gray!15}
\textbf{Reference} &
\textbf{Paradigm} &
\textbf{Input / Context} &
\textbf{Spatial Representation} &
\textbf{Sparse-Input Handling} &
\textbf{Explicit Uncertainty} &
\textbf{Distinctive Feature} \\
\midrule

\cite{teganya2022deep}
&
Completion autoencoder
&
Sparse measurements and historical radio maps
&
Regular grid
&
Rasterized
&
No
&
Learns recurring propagation structure through fully convolutional map completion
\\
\midrule

\cite{levie2021radiounet}
&
CNN / UNet
&
Building geometry and transmitter location
&
Regular 2-D grid
&
Grid-based
&
No
&
Fast environment-aware pathloss-map prediction
\\
\midrule

\cite{ma2025selfsupervised}
&
Self-supervised CNN
&
Environment maps, transmitter information, and sparse measurements
&
Regular 2-D grid
&
Grid-based
&
No
&
Improves data efficiency through masked self-supervised pretraining
\\
\midrule

\cite{jaensch2026radiomapnoisy}
&
Context-aware CNN
&
Noisy environment information and sparse observations
&
Regular 2-D grid
&
Grid-based
&
No
&
Improves robustness to errors in geometry, materials, and transmitter locations
\\
\midrule

\cite{shibli2024data}
&
Graph neural network
&
Cell information and sparse RSRP/RSRQ measurements
&
Graph
&
Graph-based
&
No
&
Models relational spatial dependencies without image-style representation
\\
\midrule

\cite{Shrestha2022DeepSC}
&
Structured hybrid learning
&
Sparse spatio-spectral measurements
&
Tensor / emitter-wise maps
&
Structured completion
&
No
&
Combines emitter decomposition with learned neural completion
\\
\midrule

\cite{chen2025actgan}
&
GAN
&
Building/Tx information and sparse observations
&
Regular 2D grid
&
Grid-based
&
No
&
Combines adversarial learning with contextual and attention modules
\\
\midrule

\cite{hu2023gan3d}
&
Unsupervised GAN
&
Incomplete measurements along constrained aircraft trajectories
&
3D grid
&
Masked 3D input
&
No
&
Reconstructs 3D radio maps without complete-map supervision
\\
\midrule

\cite{yang2025radiovae}
&
VAE / probabilistic
&
Urban topology and transmitter information
&
Regular 2D grid
&
Grid-based
&
Yes
&
Generates probabilistic radio maps with uncertainty information
\\
\midrule

\cite{ye2025adiffusion}
&
Conditional diffusion
&
Sparse measurements and environmental obstacles
&
Coarse-to-fine 2D grid
&
Conditional sparse map
&
No
&
Performs fine-grained reconstruction through two-stage diffusion
\\
\midrule

\cite{liu2026diffusion}
&
Diffusion + active sensing
&
Sparse time-aligned UAV measurements and communication state
&
Dynamic 2D grid
&
Conditional sparse map
&
Yes$^\dagger$
&
Couples communication-aware sampling, temporal alignment, and diffusion reconstruction
\\
\midrule

\cite{viet2025spatial}
&
Spatial Transformer
&
Irregular measurements and query locations
&
Gridless / continuous
&
Direct set input
&
No
&
Query-dependent attention with full spatial resolution
\\
\bottomrule

\multicolumn{7}{p{17.0cm}}{
\footnotesize
$^\dagger$Uncertainty is explicitly modeled in the Gaussian-process-based sampling and alignment stage rather than by the diffusion generator itself.
}

\end{tabular}
\end{table*}

\subsubsection{Comparative Analysis and Key Insights}
Table~\ref{tab:learning_radiomap} shows that learning-based reconstruction methods differ not only in architecture, but also in how they represent measurements, environmental context, and spatial structure. Grid-based CNN and U-Net models are effective when rasterized environmental information is available, whereas graph- and structured models better exploit relational or domain-specific structure. Generative approaches, including GANs, VAEs, and diffusion models, offer greater flexibility under severe sparsity and can model richer radio-field distributions, but generally incur higher training or inference cost. Gridless attention-based models instead operate directly on irregular measurement sets and support query-dependent prediction, although their complexity may increase with the number of observations.

Several broader lessons follow. First, no single paradigm is uniformly preferable: classical and structured approaches remain attractive in data-limited regimes, while learned models are particularly useful when propagation dependencies are difficult to specify analytically. Second, representation can be as important as model capacity; grid-based models naturally exploit image-like context, whereas graph- and attention-based approaches are better suited to irregular or geometry-dependent measurements. Third, uncertainty modeling, domain generalization, and realistic evaluation remain underdeveloped in many neural reconstruction methods. Recent high-fidelity datasets and simulation platforms are beginning to improve reproducibility and cross-model comparison \cite{jia2025urbanmimomap,liu2025openpathnet}, but broader validation across environments and sensing conditions is still needed. These limitations are especially relevant to dynamic LEO systems, where sensing geometry, propagation conditions, and measurement availability can vary rapidly.

\subsection{LEO Satellite Localization}
\label{sec:leo-localization}
\begin{figure}
    \centering
    \includegraphics[width=0.85\linewidth]{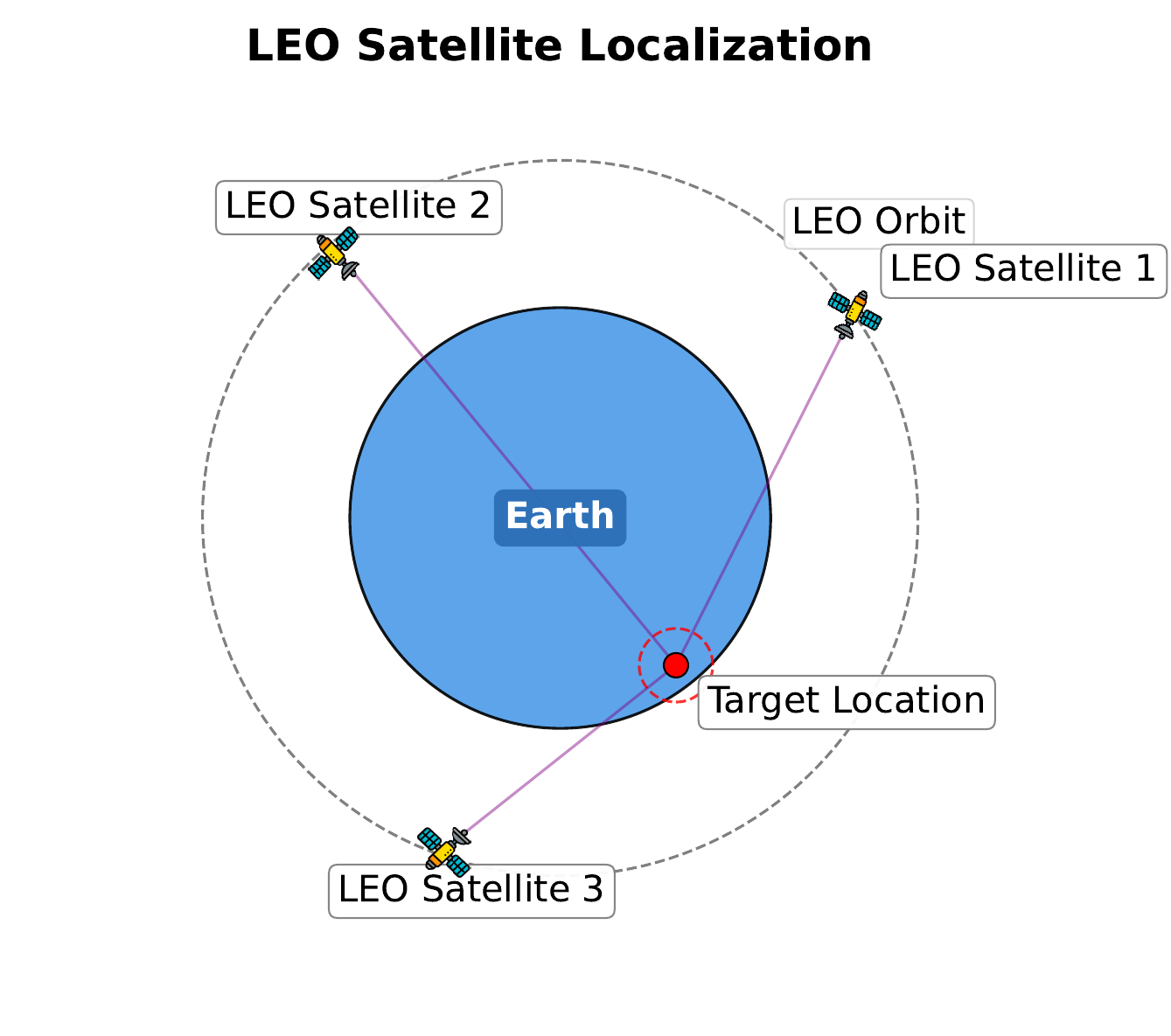}
    \caption{Illustration of a LEO satellite localization scenario, where multiple LEO satellites are used to determine the location of a ground target.}
    \label{fig:leo_example}
\end{figure}

The rapid deployment of large-scale LEO satellite constellations is reshaping global connectivity and positioning capabilities \cite{1984JBIS,handley2018delay}, as illustrated by the representative satellite--receiver geometry in Fig.~\ref{fig:leo_example}. Compared with traditional global navigation satellite systems (GNSS) \cite{hofmann2012global,bancroft2007algebraic}, LEO satellites operate at lower altitudes and higher velocities, creating new opportunities for positioning, navigation, and timing (PNT) services in non-terrestrial networks \cite{enge1994global}.

However, LEO systems differ fundamentally from conventional GNSS. Their fast motion induces strong Doppler effects \cite{guier2007satellite}, while limited synchronization leads to asynchronous measurements. The satellite–user geometry also evolves rapidly, resulting in time-varying observability and geometric dilution of precision. LEO measurements are influenced by signal conditions, orbital and clock errors, and rapidly changing satellite visibility. These effects enter positioning through different mechanisms: measurement errors affect statistical accuracy, while satellite geometry and the observation interval govern state observability \cite{torrieri2007statistical,2022leosurvey,2024leosurveystock}. We briefly summarizes LEO localization observables and model-based estimators, then surveys learning-based components for measurement extraction, temporal modeling, and multi-observation fusion. The methods are compared in terms of their use of geometry, dynamics, and uncertainty.

\subsubsection{Problem Formulation}
We consider a user with an unknown state and measurements obtained from multiple LEO satellites following nominally known trajectories.

\textit{State and Geometry.}
Let $\mathbf{x}\in\mathbb{R}^{n_x}$ denote the unknown user state. For position-centric localization we set $\mathbf{x}=\mathbf{p}\in\mathbb{R}^3$, where $\mathbf{p}$ represents the user position in the earth-centered earth-fixed (ECEF) coordinate system \cite{kaplan2017understanding}. The formulation can be extended to augmented states including velocity and clock parameters. Consider $N$ LEO satellites indexed by $\mathcal{S}=\{1,2,\dots,N\}$. The position and velocity of satellite $i$ at time $t$ are denoted by $\mathbf{s}_i(t)$ and $\dot{\mathbf{s}}_i(t)$ and are obtained from ephemeris information; the associated orbital errors may be treated as additional model uncertainty.

\textit{Generic Measurement Model.}
Let $m\in\mathcal{M}$ index an individual observation, and let $\mathcal{G}_m$ denote the associated satellite geometry, which may involve a single satellite or a satellite pair. A general measurement model is:
\begin{equation}
\mathbf{y}_m = h_m(\mathbf{x},\mathcal{G}_m) + \mathbf{n}_m,
\quad m\in\mathcal{M},
\label{eq:generic_measurement}
\end{equation}
where $h_m(\cdot)$ is a possibly nonlinear observation function and $\mathbf{n}_m$ collects measurement noise and residual modeling errors. The measurement function depends on the modality and geometry, whereas the error statistics depend on receiver, propagation, and synchronization conditions.

\textit{A Representative Estimation Objective.}
One common model-based choice is a weighted least-squares estimator:
\begin{equation}
\hat{\mathbf{x}} =
\arg\min_{\mathbf{x}}
\sum_{m\in\mathcal{M}}
\left\|
\mathbf{y}_m - h_m(\mathbf{x},\mathcal{G}_m)
\right\|^2_{\mathbf{W}_m}.
\label{eq:weighted_form}
\end{equation}
Here $\mathbf{W}_m\succeq0$ is a measurement weighting matrix. For independent, zero-mean Gaussian measurement errors with known covariances, inverse-covariance weighting corresponds to maximum likelihood estimation; other statistical assumptions lead to different estimation objectives.

\begin{figure}
    \centering
    \includegraphics[width=0.95\linewidth]{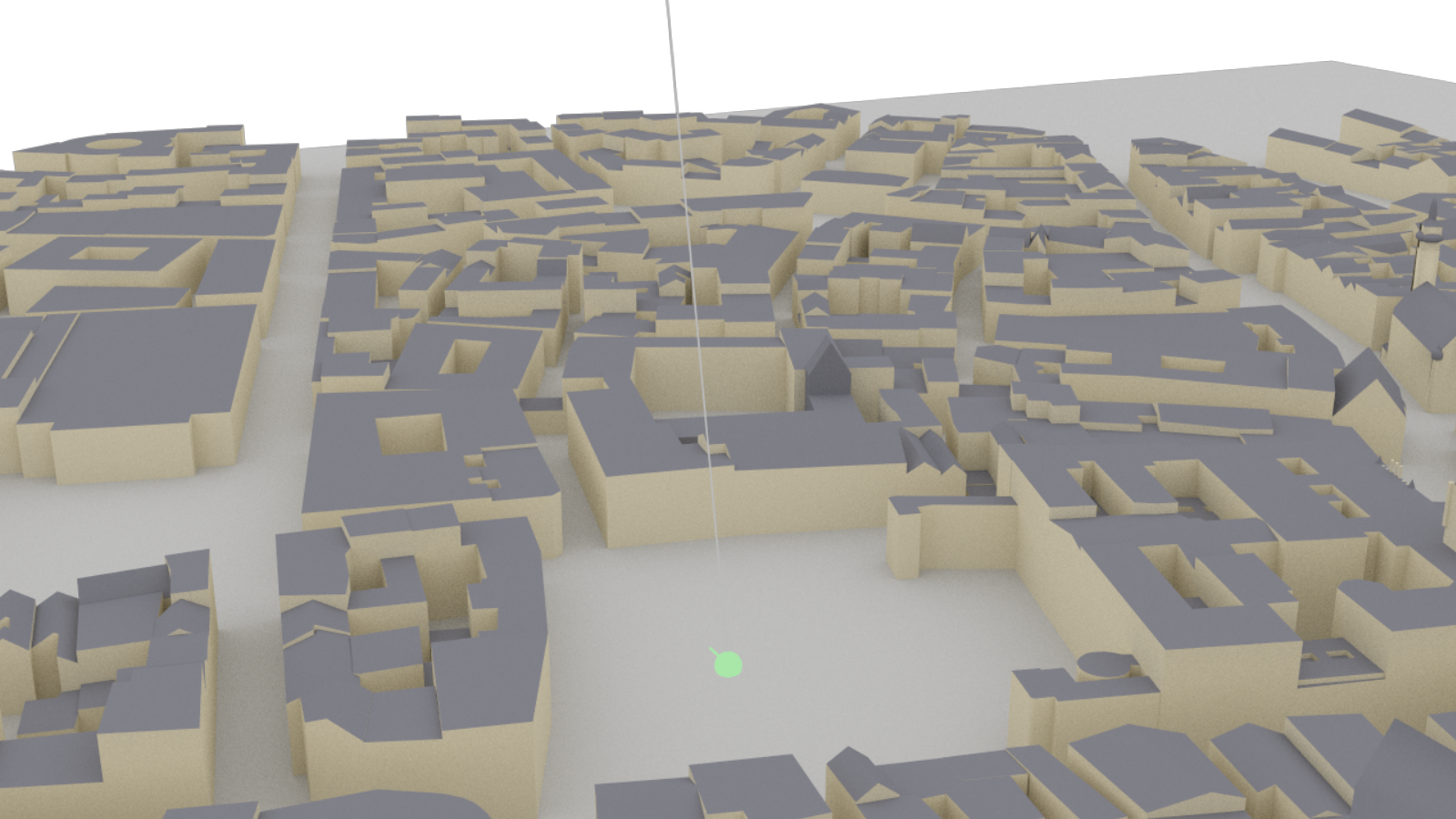}
    \caption{Sionna RT scene of the link geometry between the observer and \textit{STARLINK-11166 [DTC]} at UTC 2026-04-13 05:03:42.75. The satellite subpoint is at $(1.222477^\circ,\,104.572422^\circ)$ with an altitude of $360.890\,\mathrm{km}$, while the observer sees the satellite at $75.257^\circ$ elevation, $95.476^\circ$ azimuth, and a slant range of $372.471\,\mathrm{km}$.}
    \label{fig:sionna-leo}
\end{figure}

\textit{Problem Characteristics.}
Rapid satellite motion results in changing visibility, geometry, and observability. Signal quality, synchronization, and temporal alignment further affect the observations \cite{kaplan2017understanding}. The next subsection summarizes the main measurement modalities. Figure~\ref{fig:sionna-leo} illustrates an instantaneous LEO satellite--observer link geometry in a representative Sionna RT scene.

\subsubsection{Measurement Models}
Time-, frequency-, and power-based observations provide distinct constraints on the receiver state. The following simplified models retain their basic geometric dependence, while practical estimation may also require clock, propagation, and orbital-error terms \cite{2022leosurvey,2024leosurveystock}.

\textit{Preliminaries and Notation.}
Using the notation above, define $\mathbf{r}_i(t)\triangleq \mathbf{p}-\mathbf{s}_i(t)$, $\rho_i(t)\triangleq \|\mathbf{r}_i(t)\|$, and the LOS unit vector $\mathbf{u}_i(t)\triangleq \frac{\mathbf{r}_i(t)}{\|\mathbf{r}_i(t)\|}$.

\textit{Range / TOA and TDOA Measurements.}
Range-type observations are fundamental in satellite navigation \cite{enge1994global,del2019technical}. A simplified geometry-only model for all $i$ is:
\begin{equation}
y_i^{\mathrm{r}}(t)=\rho_i(t)+n_i^{\mathrm{r}}(t),
\label{eq:range_model}
\end{equation}
where $n_i^{\mathrm{r}}(t)$ collects residual errors in this simplified model. TOA or pseudorange observations also involve clock biases, whereas TDOA uses range differences $\rho_i(t)-\rho_j(t)$ together with any residual synchronization terms \cite{enge1994global,wang2023doppler,li2024leo}.

\textit{Doppler / FDOA-Type Measurements.}
Because LEO satellites move at high orbital velocities, Doppler observables are particularly informative for LEO-based PNT \cite{2022leosurvey,shi2023revisiting,neinavaie2021acquisition}. A standard Doppler (range-rate) model is
$y_i^{\mathrm{d}}(t)=\dot{\rho}_i(t)+n_i^{\mathrm{d}}(t)$
with
$\dot{\rho}_i(t)=\mathbf{u}_i^\top(t)(\dot{\mathbf{p}}(t)-\dot{\mathbf{s}}_i(t))$.
Thus, Doppler is determined by the projection of relative velocity onto the LOS \cite{kaplan2017understanding}. Receiver velocity and clock drift can be incorporated into an augmented state. A frequency difference of arrival (FDOA) measurement between satellites $i$ and $j$ is:
\begin{equation}
y_{ij}^{\mathrm{fd}}(t)=\dot{\rho}_i(t)-\dot{\rho}_j(t)+n_{ij}^{\mathrm{fd}}(t).
\label{eq:fdoa_model}
\end{equation}

\textit{RSS / Power-Based Measurements.}
In opportunistic LEO localization, RSS or related power metrics may also be available \cite{2024leosurveystock}. A commonly used path-loss model is:
\begin{equation}
y_i^{\mathrm{rss}}(t)=P_0-10\alpha\log_{10}\!\left(\frac{\rho_i(t)}{\rho_0}\right)+\eta_i(t),
\label{eq:rss_model}
\end{equation}
where $P_0$ is the received power at reference distance $\rho_0$, $\alpha$ is an effective path-loss exponent, and $\eta_i(t)$ captures shadowing and power fluctuations \cite{goldsmith2005wireless}. Compared with time of arrival (TOA) and Doppler, RSS provides weaker geometric sensitivity and is more environment dependent, with errors influenced by blockage, antenna patterns, beam scheduling, and multipath \cite{2022leosurvey}.

\textit{Hybrid Measurements.}
Multiple modalities can be combined within \eqref{eq:generic_measurement}; LEO observations may also be integrated with GNSS and inertial measurements \cite{2024leosurveystock,wang2023doppler}.

\textit{Measurement Comparison.}
Table~\ref{tab:mm_summary} contrasts the geometrical constraints and principal error sources of the modalities. Measurement-error statistics, geometric observability, and the information added by observations across time are related but distinct factors in localization performance.

\begin{table*}[!t]
\centering
\caption{Principal measurement modalities for LEO localization and their estimation characteristics.}
\label{tab:mm_summary}
\renewcommand{\arraystretch}{1.28}
\setlength{\tabcolsep}{5pt}
\begin{tabularx}{\textwidth}{@{}l p{3.1cm} X X@{}}
\toprule
\textbf{Modality} & \textbf{Geometric quantity} & \textbf{Geometric sensitivity} & \textbf{Representative error sources}\\
\midrule
TOA / range & $\rho_i=\|\mathbf{p}-\mathbf{s}_i\|$ & LOS diversity; clock-state coupling & Delay estimation, clock errors, multipath \cite{kaplan2017understanding,del2019technical}\\
\hline
TDOA & $\rho_i-\rho_j$ & Satellite-pair geometry & Synchronization, correlation among differenced observations \cite{ho1993solution,jardak2022potential}\\
\hline
Doppler / FDOA & $\dot{\rho}_i=\mathbf{u}_i^\top(\dot{\mathbf{p}}-\dot{\mathbf{s}}_i)$ & Radial velocity; temporal geometry evolution & Oscillator drift, Doppler-tracking errors, ephemeris errors \cite{shi2023revisiting,wang2023doppler}\\
\hline
RSS & $P_0-10\alpha\log_{10}(\rho_i/\rho_0)$ & Indirect distance dependence & Transmit power, beam pattern, blockage, shadowing \cite{goldsmith2005wireless,2024leosurveystock}\\
\bottomrule
\end{tabularx}
\end{table*}

\subsubsection{Model-Based Estimation, Geometry, and Uncertainty}
\label{subsec:uncertainty_geometry}
Classical model-based estimators relate noisy observables to receiver states using explicit measurement and, where appropriate, dynamic models. Least-squares (LS) and weighted least-squares (WLS) provide simple batch estimates, with WLS using the form in \eqref{eq:weighted_form}. Under Gaussian measurement errors with known covariance, covariance-weighted LS coincides with maximum likelihood estimation; Bayesian filters and smoothers additionally incorporate motion models and prior information \cite{van2004detection,bar2001estimation}. Robust losses or residual reweighting help limit the influence of atypical observations \cite{huber1992robust}. These methods can operate without training data and offer a physically interpretable estimation framework.

\textit{Geometry, Observability, and Uncertainty.}
Measurement-noise statistics and geometric observability have distinct effects on positioning accuracy. For additive Gaussian errors with state-independent covariance $\mathbf{R}$, the Fisher information matrix (FIM) is
\begin{equation}
\mathbf{J}(\mathbf{x})=\mathbf{H}^{\top}(\mathbf{x})\mathbf{R}^{-1}\mathbf{H}(\mathbf{x}),
\label{eq:non-linear}
\end{equation}
where $\mathbf{H}(\mathbf{x})=\partial\mathbf{h}(\mathbf{x})/\partial\mathbf{x}$ is the Jacobian. The covariance controls error scaling, while the Jacobian describes which state directions are constrained by the satellite--receiver geometry. The FIM rank and conditioning determine local observability and directional uncertainty; rapid changes in satellite visibility and line-of-sight directions make these characteristics time-dependent in LEO systems \cite{kaplan2017understanding,langley1999dilution,2022leosurvey}. Doppler-based positioning can additionally benefit from measurements collected across multiple epochs, provided receiver motion and clock evolution are appropriately modeled \cite{del2019technical,wang2023doppler}.

\textit{Two Illustrative Geometric Examples.}
The following idealized 2D range examples illustrate (i) geometric observability under equal measurement noise and (ii) the combined roles of observation precision and directional diversity.

\begin{figure}[tbp]
    \centering
    \includegraphics[width=0.95\linewidth]{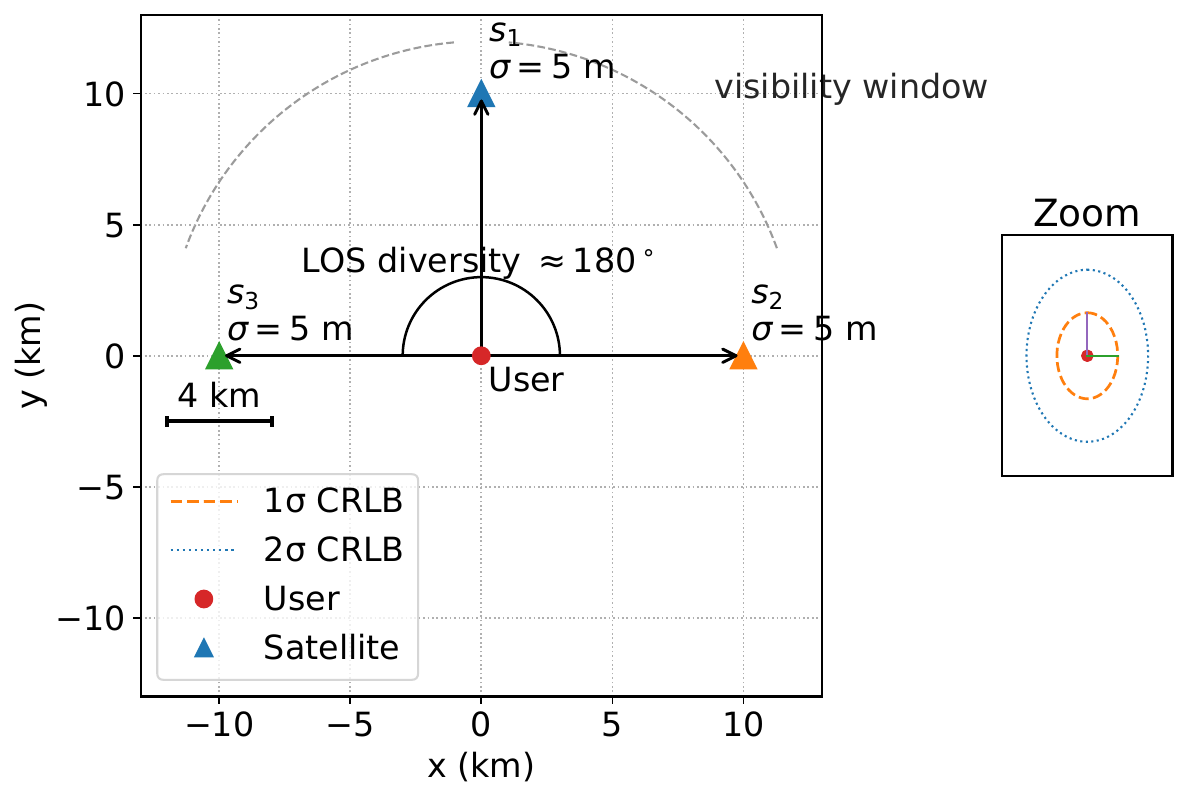}
    \caption{Illustration of 2D range localization geometry under equal measurement noise, showing geometry-dependent uncertainty and observability.}
    \label{fig:example1}
\end{figure}

\begin{example}[Geometry-Driven Anisotropy Under Equal Noise]
\label{ex:geo_heterogeneity}
Let $\mathbf{p}^{\star}=(0,0)$ and consider three satellites at $\mathbf{s}_1=(0,10)$, $\mathbf{s}_2=(10,0)$, and $\mathbf{s}_3=(-10,0)$ km. For independent range noise with $\sigma=5$ m, the corresponding line-of-sight vectors are $\mathbf{u}_1=(0,-1)$, $\mathbf{u}_2=(-1,0)$, and $\mathbf{u}_3=(1,0)$. The Fisher information matrices are:
\begin{equation}
\mathbf{J}_{123}=\frac{1}{\sigma^2}\begin{bmatrix}2&0\\0&1\end{bmatrix},\qquad
\mathbf{J}_{23}=\frac{1}{\sigma^2}\begin{bmatrix}2&0\\0&0\end{bmatrix}.
\end{equation}
With all three observations, the CRLB gives $\mathrm{std}(x)\geq\sigma/\sqrt{2}$ and $\mathrm{std}(y)\geq\sigma$. Using only $\{2,3\}$ makes the $y$ direction locally unobservable (Fig.~\ref{fig:example1}).
\end{example}

\begin{figure*}[tbp]
    \centering
    \includegraphics[width=0.9\textwidth]{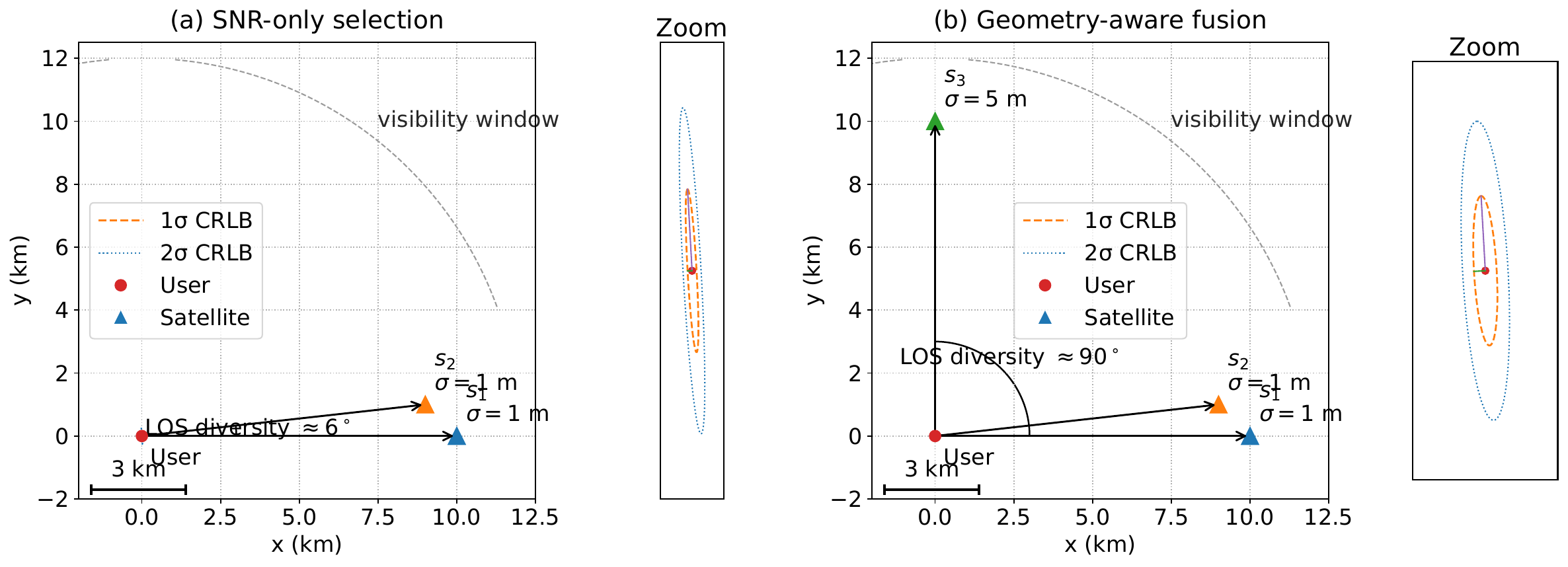}
    \caption{Illustration of 2D range localization geometry under unequal measurement noise, comparing two- and three-observation configurations.}
    \label{fig:example2}
\end{figure*}

\begin{example}[Geometric Complementarity Under Unequal Noise]
\label{ex:snr_selection_failure}
Let $\mathbf{p}^{\star}=(0,0)$ and $\mathbf{s}_1=(10,0)$, $\mathbf{s}_2=(9,1)$, and $\mathbf{s}_3=(0,10)$ km, with independent range-noise standard deviations $\sigma_1=\sigma_2=1$ m and $\sigma_3=5$ m. For $\mathbf{J}_{\mathcal S}=\sum_{i\in\mathcal S}\sigma_i^{-2}\mathbf{u}_i\mathbf{u}_i^\top$, the two higher-precision observations give:
\begin{equation}
\mathbf{J}_{12}\approx\begin{bmatrix}1.9878&0.1098\\0.1098&0.0122\end{bmatrix},\qquad
\mathrm{std}(y)\gtrsim12.77\ \mathrm{m}.
\end{equation}
Replacing observation 2 by the noisier but directionally distinct observation 3 gives $\mathbf{J}_{13}=\operatorname{diag}(1,0.04)$ and $\mathrm{std}(y)\geq5$ m for the same two-observation budget. When all three observations are combined, $\mathrm{std}(y)\gtrsim4.66$ m (Fig.~\ref{fig:example2}). Thus, noise level and directional coverage jointly determine the information available for estimating the receiver position.
\end{example}

\textit{Practical Considerations.}
The accuracy of model-based methods depends on the availability of adequate orbit and clock information, the treatment of synchronization and propagation errors, and the suitability of assumed error statistics. Measurement selection, covariance specification, robust estimation, temporal filtering, and joint multi-source estimation address different aspects of these conditions \cite{fan2024toward,2024leosurveystock}. Learning-based methods may complement these components by estimating difficult-to-model observables, systematic errors, or context-dependent fusion parameters.

\subsubsection{Learning-Based Approaches to LEO Localization}

Learning-based localization methods can be distinguished by the component they estimate, rather than by neural-network architecture alone. The direct LEO literature is still narrower than the broader GNSS learning literature. Accordingly, this review distinguishes methods tested with LEO observations from transferable methods evaluated only with GNSS data. The LEO signal-processing and navigation context is discussed in \cite{leoRefKassasAdAstra2024}, while a complementary benchmark of learning-assisted GNSS error mitigation is provided in \cite{leoRefXuGNSSReview2024}.

\textit{LEO-specific temporal modeling and model correction.}
Kozhaya \emph{et al.} compared feedforward and recurrent neural models for orbit-state propagation within a simultaneous tracking and navigation framework; their evaluation included vehicle navigation aided by real ORBCOMM observations \cite{leoRefKozhayaION2021}. El Kouba \emph{et al.} subsequently used a time-delay neural network (TDNN) to correct SGP4-propagated Starlink ephemerides and evaluated its impact on Doppler-based receiver positioning in simulation \cite{leoMLElKouba2024}. In both cases, learning supplies a model correction, while the downstream state estimate retains an explicit physical model. The distinction between experimental LEO observations and simulated downstream positioning is important when comparing their evidence.

\textit{Learned measurement screening and position correction.}
Mature GNSS work provides complementary approaches for measurements corrupted by multipath or non-line-of-sight (NLOS) reception. Hsu used signal-strength and pseudorange--Doppler consistency features for machine-learning-based measurement classification \cite{leoRefHsuITSC2017}. Zhang \emph{et al.} used long short-term memory to predict urban GNSS measurement uncertainty from temporal observation patterns \cite{leoRefZhangLSTM2021}. Kanhere \emph{et al.} instead proposed a set-based neural correction model to handle varying numbers of GNSS observations \cite{leoRefKanhereION2021}, while Xu \emph{et al.} used convolutional processing of single-differenced residual maps for urban GNSS positioning \cite{leoRefXuPositionNet2023}. These studies demonstrate different forms of learned preprocessing or correction on GNSS data; their applicability to opportunistic LEO signals requires validation with the corresponding waveforms and receiver conditions.

\textit{Hybrid optimization, contextual geometry, and learned uncertainty.}
Rather than estimating coordinates alone, some learning-assisted methods modify inputs to a model-based optimizer. Xu \emph{et al.} investigated residual-map representations for jointly addressing GNSS signal classification, correction, and selection \cite{leoRefXuGraphicalTITS2024}. A differentiable factor-graph approach learns measurement covariance parameters for smartphone GNSS positioning \cite{leoRefXuDFGOION2023}, while Hu \emph{et al.} coupled learned pseudorange-bias and weight estimates to a GNSS processing backend \cite{leoRefHuPyRTKLIB2025}. More recently, Wang \emph{et al.} used a heterogeneous graph neural network to predict covariance scale factors in GNSS/LEO precise point positioning, with real GNSS observations and simulated LEO measurements \cite{leoMLWangGraph2026}. Such models illustrate graph-based fusion and hybrid physics--learning approaches, but a learned weight or scale factor should not automatically be interpreted as a calibrated uncertainty estimate.

\textit{Methodological comparison.}
Table~\ref{tab:leo_learning_comparison} compares these studies by learned component, geometric and temporal context, uncertainty treatment, and evaluation evidence. Studies developed for GNSS are explicitly marked as transferable rather than presented as validated LEO localization systems. Set-based, convolutional, recurrent, and graph-based representations address different requirements; their reported positioning results are not directly comparable across different observation sources or evaluation protocols.

\begin{table*}[!t]
\centering
\caption{Representative learning-based methods for LEO and GNSS positioning.}
\label{tab:leo_learning_comparison}

\scriptsize
\renewcommand{\arraystretch}{1.05}
\setlength{\tabcolsep}{3.2pt}
\setlength{\aboverulesep}{0pt}
\setlength{\belowrulesep}{0pt}
\setlength{\extrarowheight}{1.5pt}

\begin{tabular}{
|>{\raggedright\arraybackslash}p{1.35cm}|
>{\raggedright\arraybackslash}p{1.55cm}|
>{\raggedright\arraybackslash}p{2.75cm}|
>{\raggedright\arraybackslash}p{1.85cm}|
>{\raggedright\arraybackslash}p{1.6cm}|
>{\raggedright\arraybackslash}p{1.72cm}|
>{\raggedright\arraybackslash}p{3.83cm}|}
\toprule
\rowcolor{gray!15}
\textbf{Reference} &
\textbf{Paradigm} &
\textbf{Measurement / Task} &
\textbf{Geometry / Context} &
\textbf{Temporal Modeling} &
\textbf{Uncertainty Handling} &
\textbf{Evaluation Setting} \\
\midrule
\cite{leoRefKozhayaION2021} &
FFNN / RNN + STAN &
LEO signal-aided navigation &
LEO orbits; inertial navigation &
RNN; tracking &
Filter covariance &
Real ORBCOMM vehicle observations (LEO) \\
\hline
\cite{leoMLElKouba2024} &
TDNN + EKF &
Starlink orbit correction; Doppler positioning &
SGP4 errors; receiver geometry &
Time-delay inputs; EKF &
EKF covariance &
Starlink ephemerides; positioning simulation (LEO) \\
\hline
\cite{leoMLWangGraph2026} &
HetGNN + FGO &
GNSS/LEO PPP covariance scaling &
Receiver--satellite graph &
Sliding-window FGO &
Learned scale factors &
Real GNSS + simulated LEO observations \\
\hline
\cite{leoRefHsuITSC2017} &
SVM classifier &
GNSS LOS / multipath / NLOS classification &
Signal strength; range--Doppler consistency &
-- &
Measurement classes &
Urban GNSS observations \\
\hline
\cite{leoRefKanhereION2021} &
Set-based DNN &
GNSS coordinate correction &
Variable-size satellite sets &
-- &
-- &
Simulated and real GNSS data \\
\hline
\cite{leoRefXuPositionNet2023} &
CNN (PositionNet) &
GNSS positioning from residual maps &
Spatial residual maps &
-- &
-- &
Real urban GNSS data \\
\hline
\cite{leoRefZhangLSTM2021} &
FCNN + LSTM &
GNSS measurement uncertainty prediction &
Observation features; urban context &
LSTM sequences &
Predicted uncertainty &
Real urban GNSS measurements \\
\hline
\cite{leoRefXuDFGOION2023} &
Diff. FGO &
GNSS covariance adaptation &
Factor graph; residuals &
Graph-state history &
Learned covariance &
Smartphone GNSS observations \\
\hline
\cite{leoRefHuPyRTKLIB2025} &
DNN + GNSS solver &
Pseudorange-bias and weight correction &
GNSS observation context &
Solver dependent &
Learned weights &
Urban GNSS data; open-source implementation \\
\bottomrule
\end{tabular}
\end{table*}

\subsubsection{Comparative Insights and Outlook}

Model-based estimators provide interpretable solutions grounded in measurement physics and satellite geometry, particularly when measurement models and error statistics are reliable. Learning-based approaches complement these methods through observable extraction, error correction, temporal modeling, and adaptive fusion. Their relative advantages depend on measurement conditions, geometric configurations, and evaluation settings. Remaining challenges include generalization across constellations and environments, uncertainty calibration, dynamic satellite geometry, and real-world LEO validation. Hybrid approaches combining learned components with geometry-constrained estimation represent a promising research direction.

\subsection{Map-Informed Resource Allocation}
\begin{figure}[!t]
    \centering
    \includegraphics[width=\linewidth]{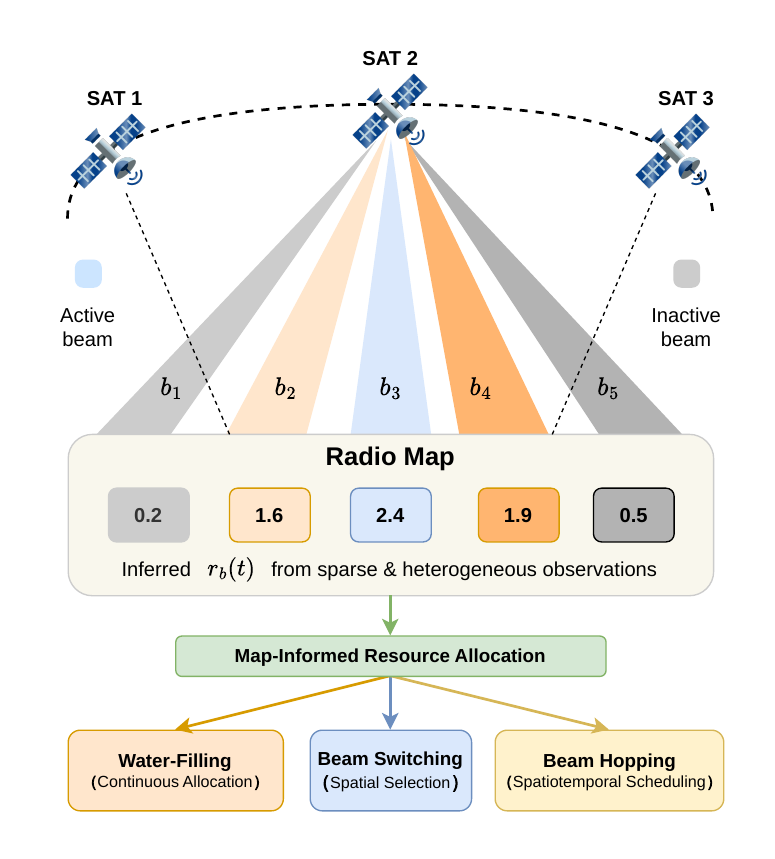}
    \caption{Illustration of map-informed resource allocation in LEO satellite networks. Beam qualities are inferred from sparse observations to form a radio map, which guides resource allocation via water-filling (continuous allocation), beam switching (spatial selection), and beam hopping (spatiotemporal scheduling).}
    \label{fig:radiomap_resource_allocation}
\end{figure}

Radio maps become operationally valuable when spatial radio knowledge is translated into network decisions. In LEO satellite networks, such decisions include power and spectrum allocation, beam selection, beam hopping, and spatiotemporal scheduling. Unlike conventional resource allocation based on instantaneous or perfectly known channel states, map-informed approaches use spatially and temporally resolved estimates of channel quality, interference, spectrum occupancy, or coverage to support decisions over locations, beams, frequencies, and time slots. As illustrated in Fig.~\ref{fig:radiomap_resource_allocation}, SC therefore acts as an information interface between sparse sensing and downstream network control.

More generally, let $\widehat{\mathcal{M}}_t$ denote the radio map or spatial radio context available at time $t$, and let $\mathbf{a}_t$ denote a resource allocation decision. Map-informed decision making can be abstractly written as:
\begin{equation}
    \mathbf{a}_t^\star
    =
    \arg\max_{\mathbf{a}_t\in\mathcal{A}_t}
    U\!\big(\mathbf{a}_t;\widehat{\mathcal{M}}_t\big),
\end{equation}
where $U(\cdot)$ denotes a network utility and $\mathcal{A}_t$ captures power, spectrum, beam, traffic, or scheduling constraints. This formulation separates the inference layer, which constructs $\widehat{\mathcal{M}}_t$, from the decision layer, which exploits the inferred spatial information.

\subsubsection{Power and Spectrum Allocation}
Power and spectrum allocation determine how limited transmission resources are distributed according to channel and interference conditions. Classical approaches such as water-filling allocate more power to favorable parallel channels according to their effective noise levels \cite{IWF2004,palomar2005practical,palomar2006tutorial}. In a spectrum-cartography setting, however, these conditions need not be directly observed; instead, radio maps provide spatially resolved estimates of channel quality, interference, and spectrum availability from distributed measurements \cite{ben2008Informed}. The allocation variables may therefore span not only frequency channels, but also beams, spatial regions, and time slots, allowing resource optimization to exploit opportunities revealed by the inferred radio environment.

An important issue is that radio maps are estimates rather than exact network states. Sparse sensing, outdated observations, and rapidly evolving interference can introduce uncertainty into downstream allocation decisions, particularly in LEO systems with fast-changing geometry and visibility. Robust formulations therefore complement map-informed optimization by accounting for adverse or uncertain interference conditions. For example, minimax formulations model transmit-power allocation and interference as coupled optimization variables \cite{ghosh2003minimax}. Along this direction, Tong et al. \cite{tongadversial2025} study adversarial water-filling, where transmit and interference powers are jointly characterized through coupled water-filling strategies. Such formulations illustrate how classical power-allocation principles can be extended when map-derived channel and interference information is uncertain or adversarial.

\subsubsection{Beam Selection and Management}
Beam selection exploits spatial variation in link quality to identify suitable transmission directions, serving beams, or satellite--user links. Classical beam-switching schemes select among a finite set of candidate beams according to received power, SINR, or achievable rate \cite{va2016beam,va2015beam}. In LEO satellite networks, however, the preferred beam can change rapidly because of satellite motion, evolving beam footprints, user location, interference, and visibility conditions. Exhaustively probing all candidate beams may therefore incur significant measurement and signaling overhead.

Radio maps provide an alternative source of spatial context by predicting channel quality or interference over candidate locations and beams. Beam selection can then exploit map-derived information to rank or filter candidate links before physical measurements or scheduling decisions are made. This interpretation is especially relevant when instantaneous observations are incomplete, as the radio map provides a broader spatial view than individual link measurements. Nevertheless, map-assisted beam management remains sensitive to prediction errors and map aging; inaccurate spatial information may lead to inappropriate beam selection, particularly near coverage boundaries or under rapidly changing blockage and interference conditions.

\subsubsection{Beam Hopping and Spatiotemporal Scheduling}
Beam hopping extends beam management from instantaneous spatial selection to joint spatial--temporal scheduling. Rather than illuminating all service regions continuously, a satellite dynamically activates subsets of beams across time according to traffic demand, channel conditions, and resource constraints \cite{hu2020dynamic,lin2022dynamic}. This mechanism is particularly attractive for LEO systems, where limited onboard power and rapidly changing coverage must be matched to spatially nonuniform and time-varying service demand.

From a spectrum-cartography perspective, dynamic radio maps can provide the spatial and temporal state information required by such scheduling decisions, including predicted channel quality, interference, coverage, and spectrum conditions across beam footprints. Beam-hopping optimization can therefore be viewed as allocating satellite resources over the spatiotemporal opportunities revealed by the radio map. Compared with static beam selection, this setting imposes stronger requirements on map freshness and temporal prediction, because scheduling decisions may become suboptimal when the inferred radio environment changes faster than the map-update cycle. These considerations motivate tighter integration between dynamic spectrum cartography, traffic prediction, and adaptive beam scheduling.

\subsubsection{Comparative Analysis and Key Insights}
The above examples illustrate that map-informed resource allocation spans fundamentally different decision spaces. Power and spectrum allocation mainly optimize continuous resource levels, beam selection performs discrete spatial decisions, while beam hopping introduces joint spatial--temporal scheduling. These mechanisms should therefore be viewed as complementary downstream uses of SC rather than direct extensions of one another. Across all three cases, the radio map acts as an intermediate representation that converts sparse observations into decision-relevant spatial knowledge. Thus, allocation performance depends not only on the optimization algorithm itself, but also on map accuracy, freshness, uncertainty, and spatial resolution. This dependence is particularly important in LEO networks, where rapidly varying geometry and interference can quickly invalidate stale map information. A key lesson is that effective map-informed control requires closer coupling between radio-map inference and downstream optimization, with uncertainty and temporal dynamics explicitly considered rather than treating the inferred map as exact ground truth.

\section{ATTENTION MECHANISMS FOR LEO SPECTRUM CARTOGRAPHY}
\label{sec:attention-leo-sc}
Attention provides a learnable mechanism for adaptive information aggregation by assigning data-dependent relevance weights to different inputs \cite{bahdanau2015neural}. From a statistical perspective, this principle is related to similarity-weighted estimators such as the Nadaraya--Watson estimator \cite{nadaraya1964estimating,watson1964smooth}, but neural attention replaces a prescribed kernel with learned compatibility functions. This enables the fusion rule to depend on the query, observation features, and surrounding context rather than on spatial proximity alone.
Such flexibility is particularly relevant to LEO spectrum cartography, where observations may be sparse, irregularly distributed, heterogeneous in modality, and strongly dependent on geometry and propagation context. In this setting, attention can support adaptive fusion for radio map reconstruction, satellite localization and state inference, and map-informed network decision making. This section first connects classical weighted aggregation with learned attention, then reviews attention and Transformer-based fusion mechanisms, their use across spectrum-cartography tasks, and gating-based extensions. We finally summarize their practical design considerations, limitations, and lessons learned, with particular emphasis on the distinction between learned relevance and calibrated measurement reliability.

\subsection{Nadaraya--Watson Estimator}
From a statistical perspective, attention-based aggregation is closely related to non-parametric regression. Given observations $\{(x_i,y_i)\}_{i=1}^{n}$, the Nadaraya--Watson (NW) estimator \cite{nadaraya1964estimating,watson1964smooth} predicts the response at a query point $x$ through similarity-weighted aggregation \cite{siraskar2024application,linke2023towards,chaudhari2021attentive}:
\begin{align}
\hat{y}(x)
&=
\sum_{i=1}^{n}\alpha(x,x_i)y_i,
\quad
\alpha(x,x_i)
=
\frac{k(x,x_i)}
{\sum_{j=1}^{n}k(x,x_j)},
\end{align}
where the normalized weights $\{\alpha(x,x_i)\}$ determine the relative contribution of each observation to the query. Here, $k(\cdot,\cdot)$ denotes a prescribed similarity kernel. A common choice is the Gaussian kernel:
\begin{align}
k(x,x_i) = \exp\Big(-\frac{\|x-x_i\|^2}{2\sigma^2}\Big),
\end{align}
where $\sigma>0$ controls the locality of the weighting.
This formulation provides a useful statistical analogy to neural attention: both construct an output through normalized, query-dependent weighted aggregation. The key distinction is that NW regression relies on a predefined kernel, whereas neural attention learns the compatibility function from data and can incorporate richer features beyond spatial proximity.

\begin{figure}[t]
    \centering
    \includegraphics[width=0.97\linewidth]{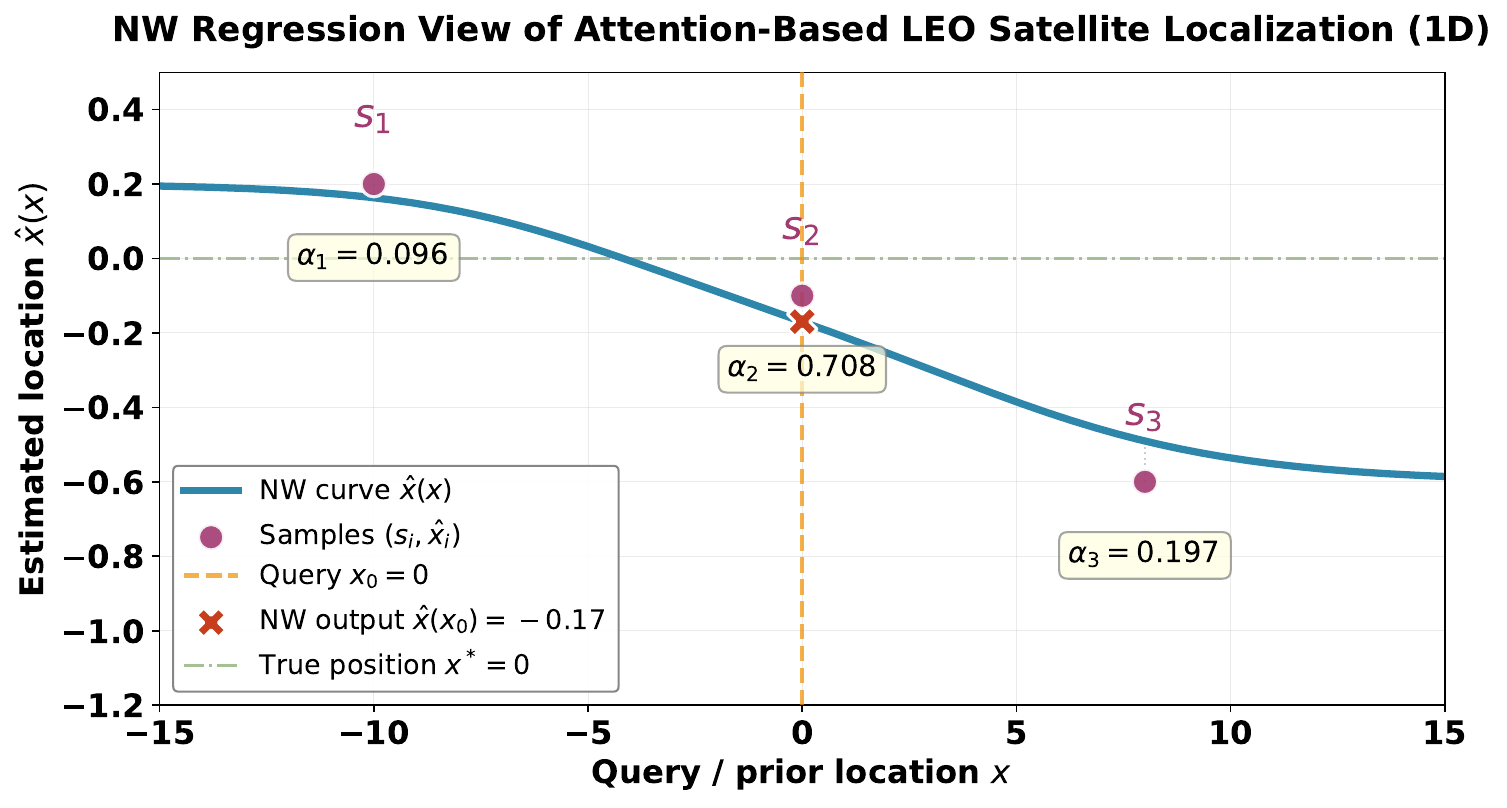}
    \caption{Nadaraya--Watson (NW) regression view of LEO satellite localization in a one-dimensional setting. Satellites at positions $s_i$ generate candidate estimates $\hat{x}_i=s_i-\rho_i$, which are fused using Gaussian-kernel similarity weights relative to a query $x_0$. The estimate $\hat{x}(x_0)\approx-0.17$ is dominated by the observation closest to the query under the selected kernel, illustrating the similarity-weighted aggregation underlying the attention analogy.}
    \label{fig:nwe-leo-localization}
\end{figure}

\begin{example}
Consider a simplified one-dimensional LEO satellite localization scenario, where a ground receiver estimates its position $x\in\mathbb{R}$ using signed pseudorange-like measurements from three visible LEO satellites. Let the true receiver position be $x^\star=0$, with satellite positions and corresponding noisy measurements $(s_1,\rho_1)=(-10,-10.2)$, $(s_2,\rho_2)=(0,0.1)$, and $(s_3,\rho_3)=(8,8.6)$. The measurements follow the simplified relation:
\begin{align}
\rho_i \approx s_i-x^\star+\varepsilon_i,
\end{align}
where $\varepsilon_i$ denotes a fixed noise realization.

Each observation induces an individual candidate position estimate $\hat{x}_i=s_i-\rho_i$, yielding $\hat{x}_1=0.2$, $\hat{x}_2=-0.1$, and $\hat{x}_3=-0.6$. To fuse these candidates, we apply the NW estimator using a coarse query location $x_0=0$:
\begin{align}
\hat{x}(x_0)
&=
\sum_{i=1}^{3}\alpha(x_0,s_i)\hat{x}_i,
\quad
\alpha(x_0,s_i)
=
\frac{k(x_0,s_i)}
{\sum_{j=1}^{3}k(x_0,s_j)}.
\end{align}
Using a Gaussian kernel with $\sigma=5$,
\begin{align}
k(x_0,s_i)
=
\exp\Big(
-\frac{(x_0-s_i)^2}{2\sigma^2}
\Big),
\end{align}
gives:
\begin{align}
k(x_0,s_1)&\approx0.135,\;
k(x_0,s_2)=1,\;
k(x_0,s_3)\approx0.278,
\end{align}
and therefore:
\begin{align}
\alpha(x_0,s_1)&\approx0.096,\;
\alpha(x_0,s_2)\approx0.708,\;
\alpha(x_0,s_3)\approx0.197.
\end{align}
The resulting estimate is:
\begin{align}
\hat{x}(x_0)
&\approx
0.096\times0.2
+
0.708\times(-0.1)
+
0.197\times(-0.6)
\notag\\
&\approx-0.17.
\end{align}

Fig.~\ref{fig:nwe-leo-localization} illustrates how NW regression performs query-dependent fusion through a prescribed similarity kernel. Satellite~2 receives the largest weight because its position is closest to the query under the selected Gaussian kernel. Importantly, these weights represent similarity-based relevance rather than calibrated measurement reliability. Neural attention extends this principle by learning the compatibility function from richer observation and contextual features.
\end{example}

\subsection{Attention in Deep Learning}
Attention can be viewed as a learnable aggregation mechanism that replaces fixed information pooling with input-dependent weighting. This perspective is particularly relevant to spectrum cartography, where observations are often sparse, irregularly distributed, and naturally represented as unordered sets rather than fixed-size tensors.

\subsubsection{Pooling as Fixed Aggregation}
Classical deep architectures commonly use deterministic aggregation operators such as mean or max pooling. Given a collection of features $\{z_i\}_{i\in\mathcal{N}}$, fixed pooling can be written as $h=\mathcal{P}\!\left(\{z_i\}_{i\in\mathcal{N}}\right)$,
where $\mathcal{P}(\cdot)$ denotes a predefined rule such as averaging or maximization \cite{krizhevsky2012imagenet}. Such operations are computationally simple and permutation invariant, but assign feature importance according to a fixed rule and therefore cannot adapt to query- or context-dependent relevance.

\subsubsection{Attention as Adaptive Aggregation}
Attention generalizes fixed pooling by introducing learnable, data-dependent weights. In its generic form, an aggregated representation can be expressed as:
\begin{equation}
z=\sum_{i=1}^{N}\alpha_i h_i,
\qquad
\alpha_i=
\frac{\exp(e_i)}
{\sum_{j=1}^{N}\exp(e_j)},
\end{equation}
where $e_i$ is a learned compatibility score that determines the relevance of feature $h_i$ to the current query or context \cite{bahdanau2015neural}. Compared with fixed pooling or predefined kernels, attention allows the aggregation rule itself to adapt to the input.

\begin{figure}[!t]
    \centering
    \includegraphics[width=0.95\linewidth]{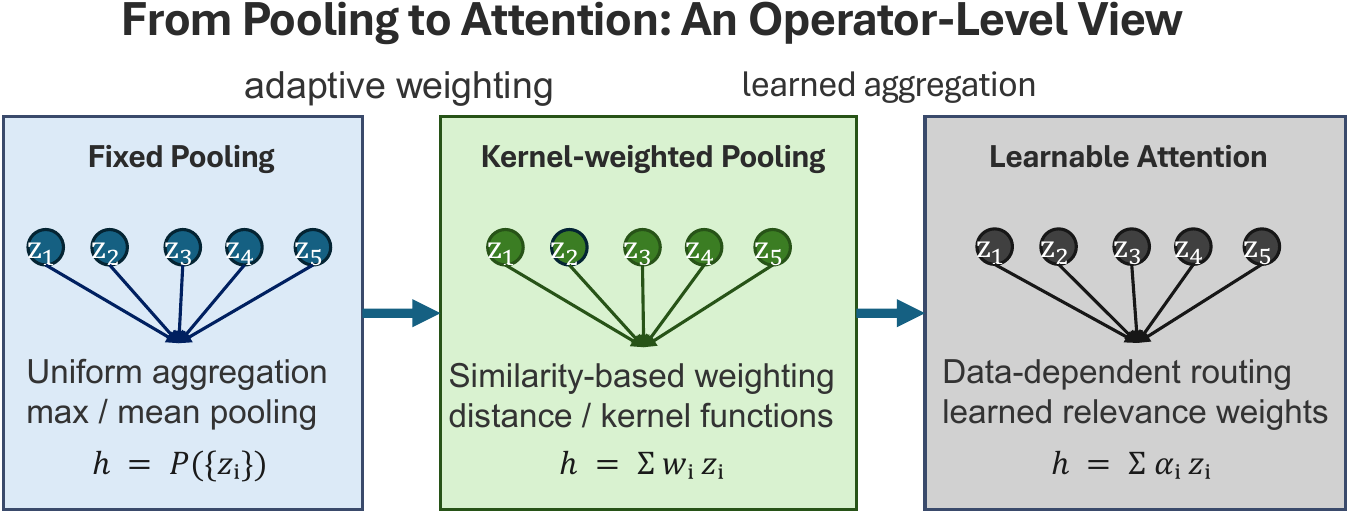}
    \caption{From pooling to attention: an operator-level view. Aggregation progresses from fixed pooling, to similarity-based weighted pooling, to learnable attention with adaptive relevance weights, reflecting a shift from static aggregation to context-dependent information fusion.}
    \label{fig:fpta}
\end{figure}

This property is particularly useful for unordered or variable-size inputs. Deep Sets established a general permutation-invariant formulation for set-valued data \cite{zaheer2017deep}, while attention-based pooling extends uniform aggregation by assigning different contributions to individual elements according to learned relevance. Related formulations have been developed for multiple-instance learning \cite{ilse2018attention} and Set Transformers \cite{lee2019set}, which use self-attention and attention-based pooling to model interactions within unordered sets. Such set-structured processing is well suited to spectrum cartography, where the number and spatial distribution of available measurements may vary across queries and time.

As illustrated in Fig.~\ref{fig:fpta}, the progression from fixed pooling to kernel weighting and learned attention reflects increasing flexibility in how observations are combined. This is particularly relevant to spectrum cartography because measurement sets may vary in size, spatial distribution, sensing modality, and geometric context. Attention provides a natural mechanism for adapting the contribution of each observation to the current query, while preserving the permutation-invariant structure required for unordered measurement sets. This aggregation perspective provides the conceptual basis for the multi-head attention mechanisms introduced next.

\subsection{Transformer Attention: Multi-Head Softmax Attention}
The Transformer architecture introduced by Vaswani et al.~\cite{vaswani2017attention} established attention as a primary mechanism for modeling interactions among input elements. In spectrum cartography, this formulation is particularly useful because measurements may interact through spatial, geometric, temporal, or modality-dependent relationships rather than through a fixed neighborhood structure.

Given an input representation $X\in\mathbb{R}^{n\times d_{\mathrm{model}}}$, Transformer attention first constructs query, key, and value representations through learned linear projections:
\begin{equation}
\mathbf{Q}=XW_Q,\qquad
\mathbf{K}=XW_K,\qquad
\mathbf{V}=XW_V,
\end{equation}
where $W_Q$, $W_K$, and $W_V$ are trainable projection matrices. The core scaled dot-product attention operation is then:
\begin{equation}
\label{eq:scaled-dot-product-attention}
\mathrm{Attention}(\mathbf{Q},\mathbf{K},\mathbf{V})
=
\mathrm{softmax}\left(
\frac{\mathbf{Q}\mathbf{K}^{\top}}{\sqrt{d_k}}
\right)\mathbf{V}.
\end{equation}
The query--key similarities determine normalized relevance weights, while the value vectors provide the information to be aggregated. Unlike fixed similarity kernels, these interactions are learned jointly with the downstream task.

Multi-head attention applies this operation in parallel across $h$ representation subspaces:
\begin{equation}
\mathbf{O}
=
\mathrm{Concat}
\left(
\mathrm{head}_1,\ldots,\mathrm{head}_h
\right)W_O,
\end{equation}
where:
\begin{equation}
\mathrm{head}_i
=
\mathrm{Attention}
\left(
\mathbf{Q}_i,\mathbf{K}_i,\mathbf{V}_i
\right).
\end{equation}
Different heads can capture complementary interaction patterns, such as spatial proximity, geometric compatibility, temporal dependence, or cross-modal relationships.

\begin{figure*}[!th]
    \centering
    \includegraphics[width=0.8\textwidth]{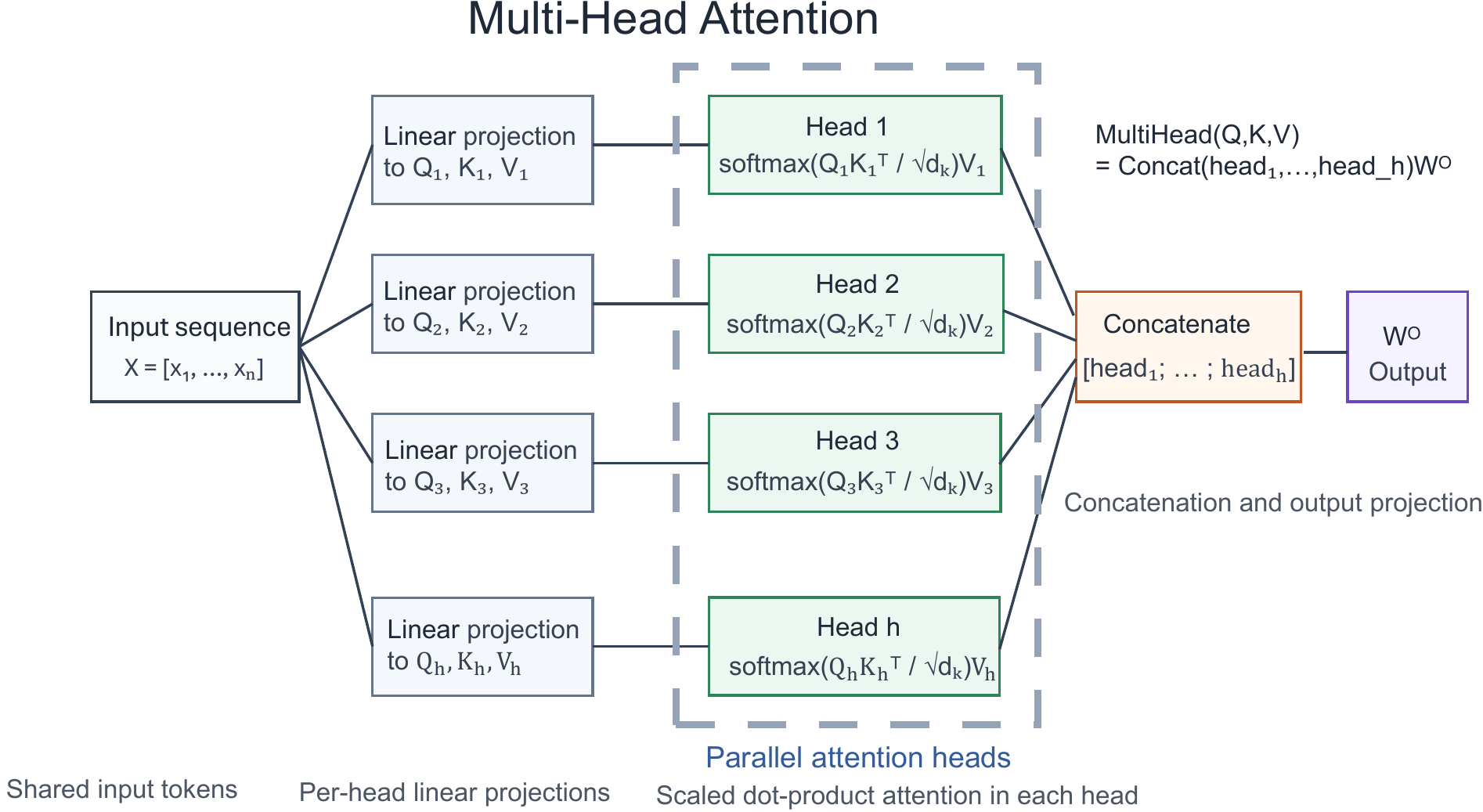}
    \caption{Multi-head attention for adaptive information fusion. The input measurements are projected into multiple query, key, and value spaces, where each attention head captures a potentially different interaction pattern before the head outputs are concatenated and linearly projected.}
    \label{fig:multi-head-attn}
\end{figure*}

For spectrum-cartography tasks, the practical role of attention depends strongly on how the input features are constructed. Measurement coordinates, signal values, satellite geometry, temporal information, environmental context, and signal-quality indicators may all be embedded into the query, key, and value representations. Attention can therefore learn context-dependent relevance across heterogeneous observations. However, the resulting attention weights should be interpreted as learned compatibility scores rather than calibrated reliability estimates unless reliability information is explicitly modeled or supervised.

\begin{example}
\label{exam:attn-leo}
Consider a simplified one-dimensional LEO localization problem with three satellite observations. The candidate position estimates are:
\begin{equation}
V=[v_1,v_2,v_3]^\top=[70,68,70]^\top.
\end{equation}
To illustrate quality-informed fusion, each observation is associated with two auxiliary features: normalized SNR and residual consistency relative to a coarse prior. The resulting key representations are:
\begin{equation}
K=
\begin{bmatrix}
0.9 & 0\\
0.4 & -2\\
0.8 & 0
\end{bmatrix},
\end{equation}
and we use the illustrative query $q=[1,1]$, which favors higher signal quality and smaller residuals.

The corresponding attention scores are:
\begin{equation}
qK^\top=[0.9,-1.6,0.8],
\end{equation}
yielding:
\begin{equation}
\alpha
=
\mathrm{softmax}(qK^\top)
\approx
[0.503,0.041,0.455].
\end{equation}
The fused position estimate is therefore:
\begin{equation}
\hat{x}_{\mathrm{attn}}
=
\sum_{i=1}^{3}\alpha_i v_i
\approx 69.9.
\end{equation}

\begin{figure}[!t]
    \centering
    \subfloat[\footnotesize Uniform pooling vs.\ attention weights]{
        \includegraphics[width=0.45\textwidth]{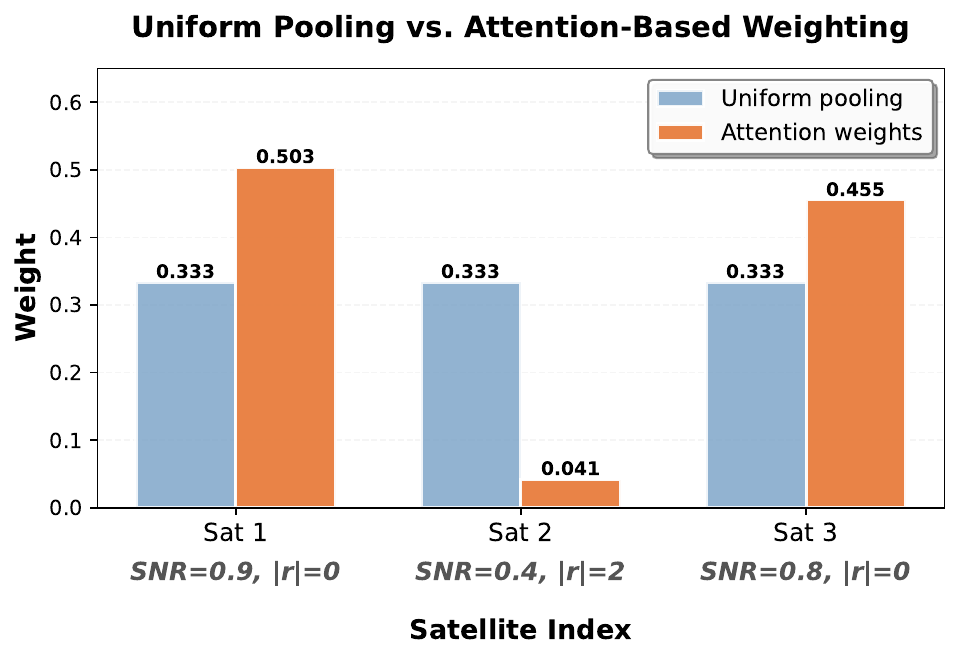}
        \label{fig:attn-leo-weights}
    }
    \\
    \subfloat[\footnotesize Candidate positions and fusion estimates]{
        \includegraphics[width=0.46\textwidth]{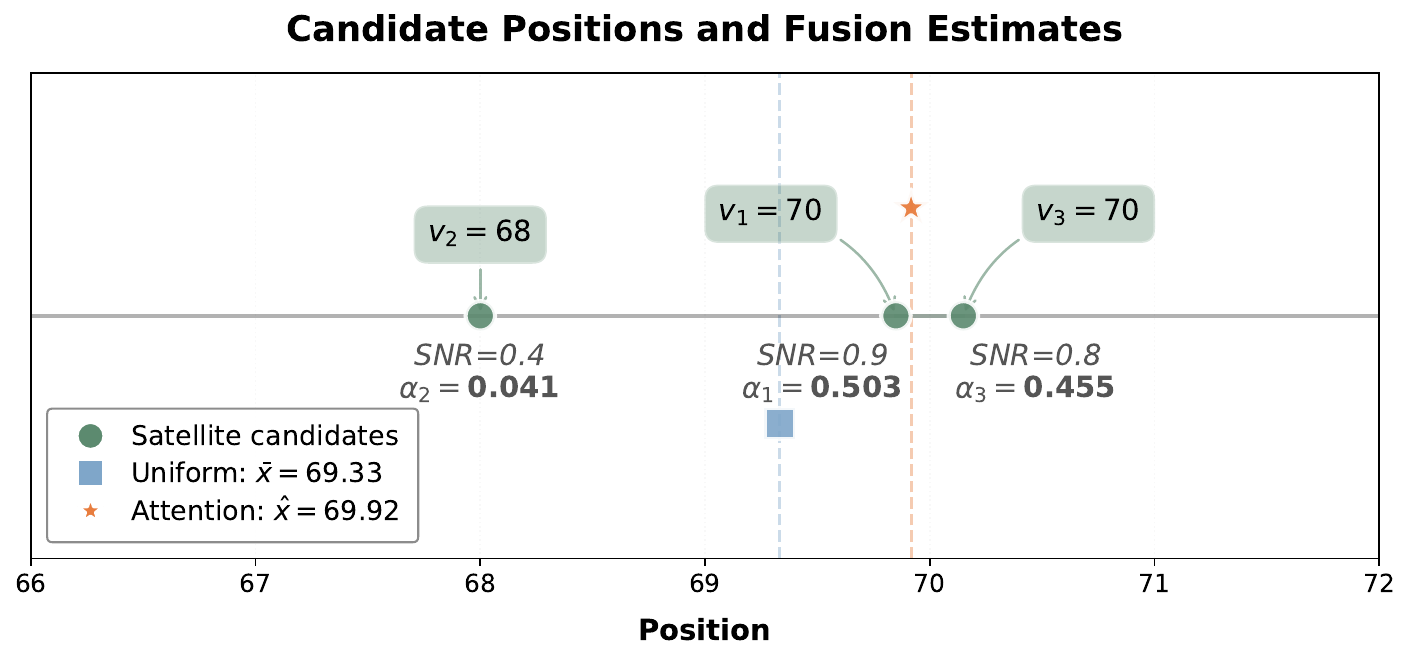}
        \label{fig:attn-leo-positions}
    }
    \caption{Illustration of feature-conditioned attention-based fusion in a one-dimensional LEO localization example. (a) Comparison between uniform pooling and attention weights when signal-quality and residual-consistency features are provided to the fusion mechanism. (b) Candidate position estimates and the corresponding fused results.}
    \label{fig:attn-leo}
\end{figure}

As shown in Fig.~\ref{fig:attn-leo}, the second observation receives a substantially smaller weight because it is less compatible with the supplied SNR and residual-consistency cues. This example illustrates how auxiliary quality-related features can influence attention-based aggregation. The resulting weights should be interpreted as feature-conditioned relevance scores rather than calibrated measurement-reliability estimates.
\end{example}

\subsection{Gated Attention}
\label{sec:gated-attention}
Standard softmax attention provides normalized relevance weighting, but its output is still determined by the learned compatibility scores and value representations. For heterogeneous radio measurements, it can be useful to introduce an additional input-dependent modulation stage that explicitly incorporates contextual, geometric, temporal, or quality-related information. Gating provides such a mechanism by adaptively controlling how much of the attention output is retained for each query or feature dimension \cite{hua2022transformer,qiu2025gated}.

\begin{figure*}[t]
    \centering
    \includegraphics[width=0.7\linewidth]{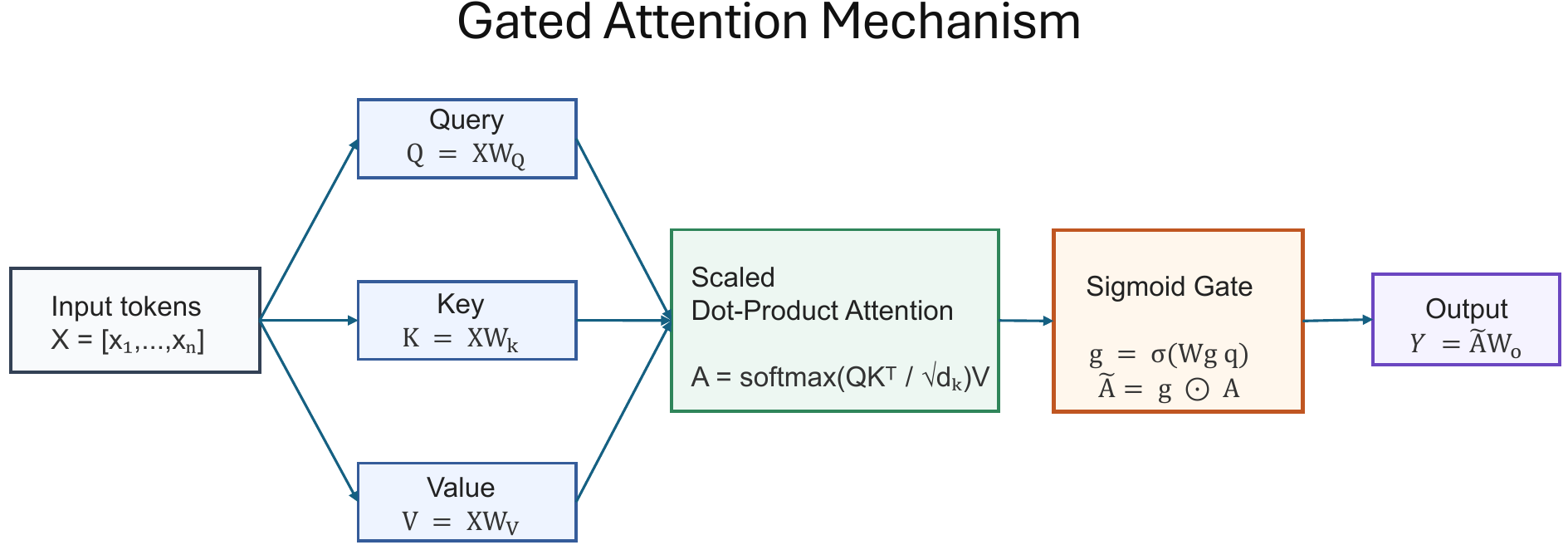}
    \caption{Simplified gated-attention mechanism for adaptive information fusion. The input features are projected into query, key, and value representations and aggregated through scaled dot-product attention. A query-dependent sigmoid gate then modulates the attention output, providing an additional input-conditioned control over information flow.}
    \label{fig:gated_attention}
\end{figure*}

Let $Y= \mathrm{softmax}\big(\frac{QK^\top}{\sqrt{d_k}}\big)V$ denote the standard attention output. As illustrated in Fig.~\ref{fig:gated_attention}, a gated representation can be written as:
\begin{equation}
\label{eq:gated-attention}
Y' = Y\odot\sigma(XW_g),
\end{equation}
where $X$ denotes the hidden or contextual representation used to generate the gate, $W_g$ is a learnable projection matrix, $\sigma(\cdot)$ is typically a sigmoid function, and $\odot$ denotes element-wise multiplication. The gating term therefore acts as an additional input-dependent filter on top of softmax attention, allowing the model to selectively modulate the information passed to subsequent layers.

For SC tasks, the usefulness of gating depends strongly on the information used to construct the gate. Signal-quality indicators, satellite--user geometry, temporal context, environmental features, residual consistency, or uncertainty estimates can all be incorporated into the gating representation. In this way, gating can complement attention by introducing an additional mechanism for feature-conditioned modulation beyond normalized query--key similarity.

\begin{example}
\label{exam:g-attn-leo}
We revisit Example~\ref{exam:attn-leo}, where the attention weights are $\alpha\approx[0.503,0.041,0.455]$.
To illustrate an additional gating stage, suppose that each observation is assigned a scalar gate derived from the same quality-related features:
\begin{equation}
g_i=\sigma(\mathbf{x}_i^\top w),
\qquad
w=[2,1]^\top.
\end{equation}
Using:
\begin{equation}
X=
\begin{bmatrix}
0.9 & 0\\
0.4 & -2\\
0.8 & 0
\end{bmatrix},
\end{equation}
the resulting gate values are approximately:
\begin{equation}
g\approx[0.858,0.231,0.832].
\end{equation}

The gated aggregation coefficients are therefore proportional to $\alpha_i g_i$, so the second observation is further attenuated because it has both a small attention weight and a small gate value.
This example illustrates the complementary roles of attention and gating: attention determines query-dependent relevance across observations, whereas gating provides an additional feature-conditioned modulation stage. The gate values should be interpreted as learned control signals rather than calibrated reliability probabilities.
\end{example}

\subsection{Software and Implementation Frameworks}
Attention-based spectrum-cartography models can be implemented using mainstream deep-learning frameworks such as PyTorch, TensorFlow, and JAX. PyTorch provides flexible eager execution, automatic differentiation, and mature support for custom attention modules and distributed training \cite{paszke2019pytorch,paszke2017automatic,ansel2024pytorch}. TensorFlow offers a production-oriented ecosystem with automatic differentiation, graph execution, distributed training, and deployment support across heterogeneous devices \cite{abadi2016tensorflow,shukla2018machine}. JAX combines a NumPy-like interface with composable program transformations such as automatic differentiation, just-in-time compilation, vectorization, and accelerator execution through XLA, making it attractive for high-performance attention and Transformer workloads \cite{bradbury2018jax,frostig2019compiling}.

For spectrum-cartography applications, the choice of framework is typically determined less by the attention mechanism itself than by practical considerations such as hardware support, distributed execution, deployment environment, and integration with simulation or communication-system toolchains. PyTorch is widely adopted for research prototyping, TensorFlow provides mature deployment and monitoring tools, while JAX is particularly suitable for compiler-oriented and highly vectorized implementations. In all cases, efficient implementation becomes increasingly important as the number of measurements, attention heads, spatial queries, or temporal steps grows.

\subsection{Capabilities and Limitations of Attention Mechanisms}

\textit{Capabilities.}
Attention mechanisms enable learnable, context-dependent aggregation by adaptively weighting inputs according to their relevance to a query \cite{lee2019attention,chaudhari2021attentive}. Unlike fixed pooling or predefined similarity kernels, attention captures interactions among heterogeneous and irregular observations, making it particularly suitable for spectrum cartography, where measurement relevance depends on spatial location, satellite geometry, temporal context, sensing modality, and propagation conditions. Self-attention models interactions among input elements, while cross-attention supports multi-modal fusion and query-conditioned inference. Multi-head attention further captures complementary interaction patterns in parallel \cite{guo2022attention,brauwers2023general}. These mechanisms can be incorporated into convolutional, graph-based, set-based, and Transformer architectures.
In LEO spectrum cartography, these capabilities facilitate query-conditioned radio map reconstruction, multi-satellite and multi-modal information fusion, and context-aware decision making. By dynamically adapting information aggregation to varying observation conditions, attention provides a flexible mechanism for handling heterogeneous measurements and complex spatio-temporal dependencies.

\vspace{1mm}
\textit{Limitations.}
Standard self-attention incurs quadratic time and memory complexity due to pairwise query--key interactions, posing scalability challenges for large measurement sets, dense spatial queries, long temporal histories, and multi-satellite observations \cite{shen2021efficient,keles2023computational}. Although sparse, localized, low-rank, and hierarchical variants can reduce computational costs, they may limit global interactions or introduce additional assumptions. Moreover, attention provides weaker task-specific inductive biases than models explicitly incorporating spatial locality, graph topology, or physical constraints \cite{edelman2022inductive}, potentially reducing generalization under limited training data. Incorporating geometric features, environmental context, and physics-based priors can improve robustness and transferability. Another limitation is that attention weights reflect learned relevance rather than calibrated measurement reliability or uncertainty, motivating explicit quality modeling and probabilistic fusion, particularly under rapidly changing LEO conditions. Finally, memory, energy, and latency constraints on resource-limited platforms motivate efficient attention architectures, model compression, and collaborative ground--satellite processing for practical deployment.

\subsection{Attention-Based LEO Localization and State Inference}
\label{sec:attn-leo-localization}
Recent work has explored attention-based learning for LEO localization and related state-inference tasks, including user positioning, orbit prediction, clock-bias estimation, interference monitoring, and anomaly detection. Across these applications, attention is primarily used to model context-dependent interactions among heterogeneous observations, temporal states, or multiple information sources. Rather than treating attention as an explicit estimator of measurement reliability, it is more appropriately viewed as a learned fusion mechanism whose weights depend on the task, representation, and available contextual features. Table~\ref{tab:attn-leo-summary} summarizes representative approaches.

\subsubsection{Unified Modeling Perspective}
Attention-based LEO state inference can be viewed as context-dependent fusion of heterogeneous observations. Let $\mathbf{z}_i$ denote the feature representation associated with the $i$-th observation and let $\mathbf{q}$ denote a query or contextual representation. A generic attention-based fusion rule can be written as:
\begin{equation}
\mathbf{u}
=
\sum_{i=1}^{M}
\alpha_i(\mathbf{q},\mathbf{z}_i)\mathbf{v}_i,
\qquad
\alpha_i
=
\frac{
\exp\!\left(g(\mathbf{q},\mathbf{z}_i)\right)
}{
\sum_{j=1}^{M}
\exp\!\left(g(\mathbf{q},\mathbf{z}_j)\right)
},
\label{eq:attn-leo-fusion}
\end{equation}
where $\mathbf{v}_i$ denotes the value representation and $g(\cdot,\cdot)$ is a learned compatibility function. The fused representation $\mathbf{u}$ is then mapped to the desired state estimate through a task-specific estimator:
\begin{equation}
\hat{\mathbf{x}}
=
f_{\theta}(\mathbf{u},\mathbf{c}),
\label{eq:attn-leo-state}
\end{equation}
where $\mathbf{c}$ may include geometric, temporal, environmental, or model-based state information.

This formulation covers multi-satellite fusion, temporal attention, multimodal integration, and residual correction without requiring the attention weights themselves to represent measurement reliability. When signal-quality, geometric, or uncertainty-related features are explicitly included in the observation representation, attention may learn to exploit such cues during inference; however, the resulting weights remain task-dependent relevance scores rather than calibrated confidence measures.

\begin{table*}[!t]
\centering
\caption{Representative attention-based methods for LEO localization and state inference.}
\label{tab:attn-leo-summary}

\setlength{\tabcolsep}{4pt}
\setlength{\extrarowheight}{1.5pt}
\renewcommand{\arraystretch}{0.9}

\begin{tabular}{
|>{\raggedright\arraybackslash}p{1.8cm}
|>{\raggedright\arraybackslash}p{2.4cm}
|>{\raggedright\arraybackslash}p{4.0cm}
|>{\raggedright\arraybackslash}p{4.0cm}
|>{\raggedright\arraybackslash}p{4.0cm}|
}
\hline
\rowcolor{gray!15}
\textbf{Work} &
\textbf{Inference Target} &
\textbf{Input / Context} &
\textbf{Role of Attention} &
\textbf{Remaining Issue} \\
\hline

\cite{zhao2024new}
&
NLOS/LOS state
&
Pseudorange, Doppler, geometry, and environmental features
&
Transformer-based multimodal fusion
&
Robustness to noise and synchronization errors
\\
\hline

\cite{huang2024global}
&
Orbit trajectory
&
Multi-scale orbital dynamics
&
Global--local temporal attention
&
Interpretability and uncertainty modeling
\\
\hline

\cite{yaman2025adaptive}
&
User position
&
Environment-dependent positioning information
&
Adaptive model or feature weighting
&
Cross-environment generalization
\\
\hline

\cite{cai2024attention}
&
Interference state
&
Spatio-temporal monitoring information
&
Multi-head global--local fusion
&
Multi-source latency and scalability
\\
\hline

\cite{kumar2024deepsatloc}
&
User position
&
GNSS, LEO, and inertial information
&
Attention-assisted GNSS correction and hybrid fusion
&
Performance under prolonged GNSS degradation
\\
\hline

\cite{cai2024enhancing}
&
Clock bias
&
Historical clock-bias sequence
&
Attention-augmented temporal modeling
&
Long-term non-stationarity
\\
\hline

\cite{re2024transformers}
&
Orbit anomaly
&
Orbital residual time series
&
Transformer-based residual modeling
&
Onboard computational cost
\\
\hline

\cite{yuan2024research}
&
Orbit prediction
&
Physical orbit model and residual sequence
&
Attention-based error refinement
&
Complexity--accuracy trade-off
\\
\hline

\end{tabular}
\end{table*}

\subsubsection{Representative Advances}
Existing studies (summarized in Table~\ref{tab:attn-leo-summary}) reveal three recurring roles of attention in LEO localization and state inference. The first is heterogeneous observation fusion. Attention-based models can combine measurements with different modalities, geometries, and temporal characteristics by learning context-dependent interactions among features. For example, Transformer-based models have been used to combine pseudorange, Doppler, satellite geometry, and environmental information for NLOS-related inference \cite{zhao2024new}, while DeepSatLoc integrates attention-assisted GNSS correction with LEO and inertial information in a hybrid localization framework \cite{kumar2024deepsatloc}.

The second role is temporal and multi-scale modeling. LEO systems exhibit rapidly evolving geometry, orbital perturbations, clock drift, and other non-stationary dynamics. Global--local attention and Transformer-based residual models have therefore been explored for orbit prediction, anomaly detection, and clock-bias estimation \cite{huang2024global,re2024transformers,cai2024enhancing,yuan2024research}. In these settings, attention primarily captures dependencies across time scales rather than directly estimating measurement quality.

A third role is adaptive fusion under changing operating conditions. Attention can modulate the contribution of different measurements, features, or models according to the current context \cite{yaman2025adaptive,cai2024attention}. When quality- or geometry-related features are explicitly provided, such mechanisms can exploit them during estimation. However, attention weights should be interpreted as task-dependent relevance scores rather than probabilistic measures of reliability.

\subsubsection{Comparative Analysis and Key Insights}
The surveyed studies indicate that attention is particularly beneficial for LEO state inference involving heterogeneous observations, long temporal dependencies, and context-dependent interactions. Rather than replacing classical estimation, attention complements model-based methods by enabling adaptive feature extraction and fusion while preserving geometric, dynamical, and uncertainty structures. Nevertheless, several challenges remain. Attention weights lack calibrated probabilistic interpretations and should not be equated with measurement reliability or confidence. Generalization across constellations, propagation environments, and operating conditions remains insufficiently explored, while physical constraints and uncertainty propagation are often weakly integrated into learned models. Moreover, computational and data demands may hinder onboard deployment, motivating lightweight attention mechanisms and hybrid model--learning architectures for practical LEO systems.

\subsection{Representative Case: Attention-Assisted LEO Localization}
\label{sec:attn-leo-case}
DeepSatLoc~\cite{kumar2024deepsatloc} provides a representative example of how learned attention can be combined with model-based state estimation. Rather than replacing the localization pipeline with an end-to-end neural network, the framework uses an attention-based module to correct degraded GNSS information and then integrates the corrected observations with LEO and inertial measurements through an extended Kalman filter (EKF). This hybrid design illustrates a broader role of attention in LEO state inference: learning observation-dependent corrections while retaining explicit dynamical and geometric structure in the estimator.

\subsubsection{Problem Setting}
In urban environments, GNSS pseudorange measurements can be severely affected by blockage, multipath, and NLOS propagation. A representative measurement model is:
\begin{equation}
p_k
=
\|\mathbf{r}_u-\mathbf{r}_{k}\|
+
c\Delta t
+
I_k
+
T_k
+
\epsilon_{\mathrm{NLOS},k}
+
\epsilon_k,
\label{eq:deepsatloc-pseudorange}
\end{equation}
where $\mathbf{r}_u$ is the user position, $\mathbf{r}_{k}$ is the satellite position, $c\Delta t$ denotes the receiver clock term, $I_k$ and $T_k$ represent ionospheric and tropospheric effects, and $\epsilon_{\mathrm{NLOS},k}$ and $\epsilon_k$ capture NLOS bias and measurement noise. Under severe blockage or multipath, GNSS-only positioning may become unreliable, while inertial dead reckoning accumulates drift. LEO-derived Doppler and angle-related measurements provide complementary geometric information that can support localization under degraded GNSS conditions.

\subsubsection{Attention-Assisted Correction and Hybrid Fusion}

DeepSatLoc uses an attention-based correction module to process satellite-dependent and temporally varying GNSS features. For each visible satellite, the input may include signal-quality indicators, residual information, geometry-related features, and prior state information. Spatial attention models interactions across satellites, while temporal attention captures dependencies across successive epochs.

Let $\mathcal{G}_t$ denote the observation window available at time $t$. The learned correction module produces a GNSS position correction:
\begin{equation}
\Delta \mathbf{r}_u^t
=
f_{\theta}(\mathcal{G}_t),
\qquad
\hat{\mathbf{r}}_u^t
=
\mathbf{r}_{\mathrm{GNSS}}^t
+
\Delta \mathbf{r}_u^t,
\label{eq:deepsatloc-correction}
\end{equation}
where $f_{\theta}$ denotes the attention-based correction network.

The corrected GNSS estimate is subsequently incorporated into an EKF together with inertial propagation and LEO-derived measurements. This separation between learned observation correction and model-based state estimation is attractive because the neural component handles complex data-dependent biases, while the EKF retains explicit temporal dynamics and covariance propagation. In this hybrid architecture, attention is therefore used to improve the representation and correction of degraded observations rather than to replace the underlying state estimator.

\subsubsection{Insights and Limitations}
DeepSatLoc demonstrates the benefits of combining attention-based learning with model-based estimation for LEO-assisted localization. Attention enables adaptive feature extraction and fusion, while the EKF maintains explicit state dynamics and uncertainty propagation. However, learned corrections may generalize poorly across different propagation environments, receiver hardware, and satellite configurations. Moreover, attention weights do not provide calibrated observation covariance, complicating their integration with Bayesian filters. Doppler and angular measurements may also be affected by blockage, synchronization errors, and limited satellite visibility. These challenges motivate closer integration of adaptive learning, explicit uncertainty modeling, and geometry-aware state estimation.

\section{ILLUSTRATIVE ATTENTION-BASED INFERENCE FOR LEO SPECTRUM CARTOGRAPHY}
\label{sec:attention-leo-rem-model-method-simu}
Building on the attention mechanisms and representative applications reviewed in the preceding sections, this section provides two concise illustrative examples to clarify how similarity-based and learnable attention can be instantiated in LEO spectrum-cartography tasks. The examples consider satellite localization and radio map reconstruction, respectively, and are intended to highlight the transition from fixed weighting rules to query-dependent aggregation.
These examples are not proposed as new state-of-the-art algorithms or comprehensive performance benchmarks. Instead, they serve as pedagogical demonstrations of how attention can be integrated with task-specific measurement models and how its behavior differs from classical fixed or kernel-based aggregation. Particular emphasis is placed on the interpretation of learned relevance, the role of contextual information, and the limitations of conclusions drawn from simplified synthetic settings.

\subsection{Illustrative Kernel-Based Temporal
Weighting for LEO Localization}
We consider LEO satellite localization from a time-ordered sequence of ground observations, $\mathcal{D}=\{x_i\}_{i=1}^{n}$, where each $x_i$ may contain Doppler, angular, or timing measurements collected during a satellite pass. To illustrate the connection between classical kernel smoothing and attention-style aggregation, we consider a Nadaraya--Watson kernel-based weighting mechanism for aggregating localization hypotheses derived from sequential observations.

\subsubsection{NW Kernel Weighting Formulation}
For each index $i$, the normalized kernel weights are defined as:
\begin{equation}
\hat{K}_{ij}=\frac{K_{ij}}{\sum_{j'=1}^{n}K_{ij'}}.
\end{equation}
Here, $K_{ij}$ is a positive kernel that characterizes similarity between observations, while the bandwidth parameter $h$ controls the effective weighting neighborhood. In the numerical illustration below, the kernel depends on temporal proximity, providing a classical similarity-weighted analogue of attention rather than a fully learnable attention mechanism.

\begin{figure*}[!t]
\centering
\subfloat[]{\includegraphics[width=0.47\linewidth]{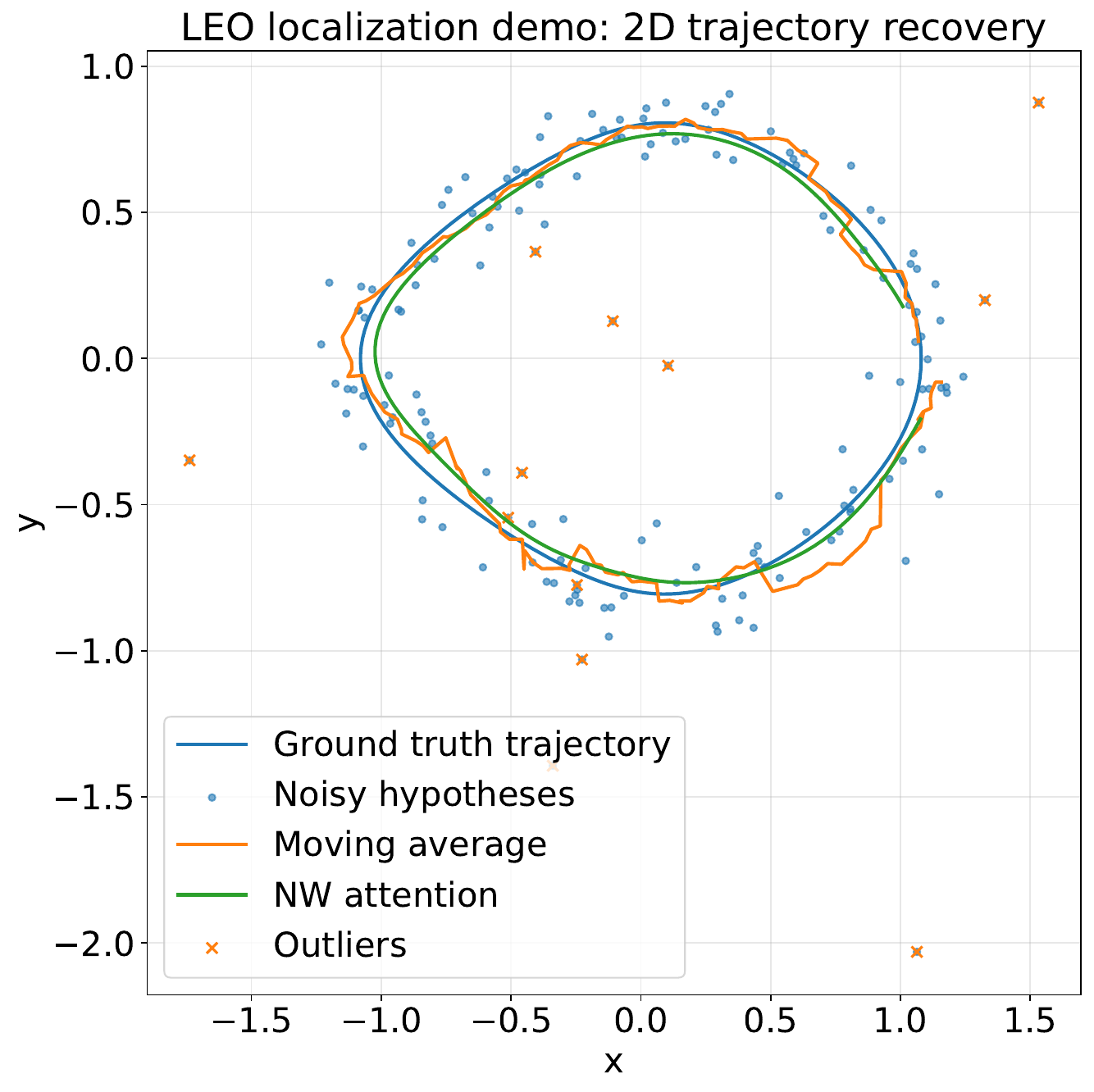}}
\hfill
\subfloat[]{\includegraphics[width=0.47\linewidth]{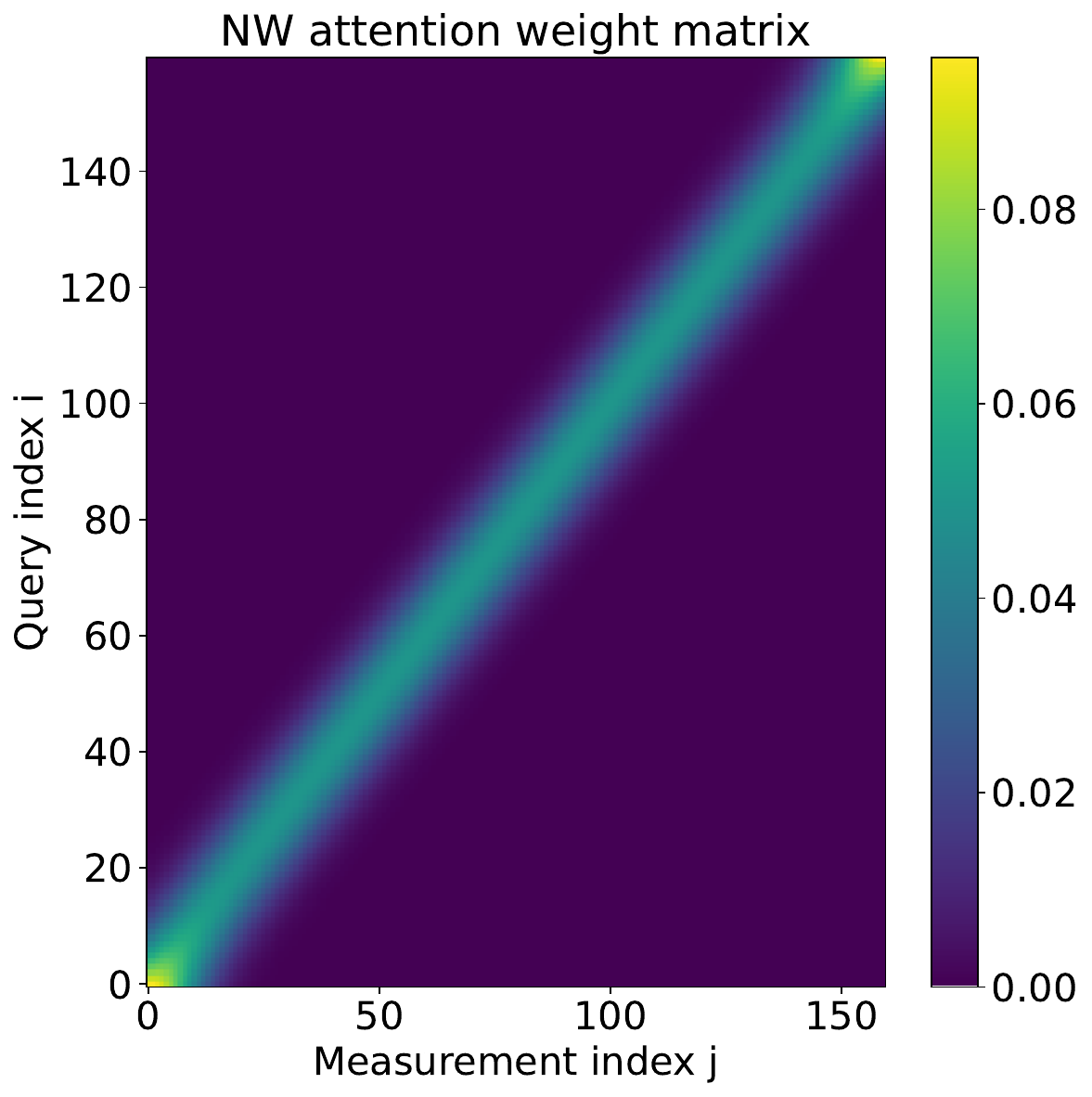}}
\caption{Illustration of kernel-based temporal weighting for synthetic LEO satellite localization. (a) Two-dimensional trajectory reconstruction from noisy localization hypotheses. (b) Normalized kernel weights $\hat{K}_{ij}$, showing the temporal locality induced by the Gaussian kernel.}
\label{fig:leo_nw}
\end{figure*}

\subsubsection{State Estimation}
Let $z_j$ denote a localization hypothesis derived from measurement $x_j$. The NW kernel-based estimate at time index $i$ is given by:
\begin{equation}
\hat{y}_i=\sum_{j=1}^{n}\hat{K}_{ij}z_j.
\end{equation}
This estimator performs normalized weighted aggregation of localization hypotheses, with the relative contributions determined by the kernel similarities. For a temporally localized kernel, observations closer in time receive greater weights, illustrating the role of locality in attention-style aggregation.

\subsubsection{Numerical Illustration}
We illustrate the kernel-based weighting mechanism using a synthetic two-dimensional trajectory generated over observation times $\{t_i\}_{i=1}^{n}$:
\begin{equation}
y_i^\star =
\begin{bmatrix}
\cos(2\pi t_i) + 0.08\cos(6\pi t_i) \\
0.8\sin(2\pi t_i) + 0.05\sin(4\pi t_i)
\end{bmatrix}.
\end{equation}
Noisy localization hypotheses are formed as $z_i=y_i^\star+\eta_i$, where $\eta_i$ is Gaussian noise with standard deviation $\sigma=0.1$, and $7.5\%$ of observations are further corrupted as outliers. We compare three estimators: raw hypotheses, a moving-average (MA) smoother, and the NW kernel smoother with Gaussian kernel weights:
\begin{equation}
K_{ij}=\exp\!\left(-\frac{(t_i-t_j)^2}{2\sigma^2}\right),
\end{equation}
where $\sigma$ controls the temporal weighting span.

For a representative run with $n=160$, the RMSE values are $\text{RMSE}_{\text{raw}}=0.265$, $\text{RMSE}_{\text{MA}}=0.064$, and $\text{RMSE}_{\text{NW}}=0.068$. Both smoothing methods reduce the estimation error relative to the raw hypotheses, with MA achieving a slightly lower RMSE in this run. As shown in Fig.~\ref{fig:leo_nw}, the normalized kernel-weight matrix exhibits a banded structure along the diagonal, reflecting the temporal locality imposed by the Gaussian kernel. This synthetic illustration highlights the connection between classical kernel-based weighting and the attention perspective in sequential estimation, without drawing broader conclusions about LEO localization performance.

\subsection{Illustrative Attention-Based Radio Map Reconstruction}
This subsection provides a simple example of how learnable attention can be used for query-conditioned radio map reconstruction. The purpose is to contrast learned aggregation with fixed distance-based or kernel-based interpolation, rather than to establish a general performance advantage over existing reconstruction methods.

\subsubsection{Query-Conditioned Formulation}
Consider a sparse set of radio measurements $\mathcal{D}=\left\{(\mathbf{x}_i,y_i)\right\}_{i=1}^{N}$, where $\mathbf{x}_i\in\Omega\subset\mathbb{R}^{2}$ denotes the measurement location and $y_i = g(\mathbf{x}_i)+\varepsilon_i$ is the observed radio quantity associated with an unknown spatial field $g(\cdot)$.
For a query location $\mathbf{q}\in\Omega$, instead of prescribing the contribution of each observation according to distance or a fixed kernel, we use a learnable scoring function:
\begin{equation}
e_i(\mathbf{q})
=
a_{\theta}
\left(
[\mathbf{q},\mathbf{x}_i,y_i]
\right),
\end{equation}
where $a_{\theta}(\cdot)$ denotes a parameterized compatibility model. The corresponding attention weights are:
\begin{equation}
\alpha_i(\mathbf{q})
=
\frac{
\exp\!\left(e_i(\mathbf{q})\right)
}{
\sum_{j=1}^{N}
\exp\!\left(e_j(\mathbf{q})\right)
}.
\end{equation}
The radio field at the query location is then predicted as:
\begin{equation}
\hat g(\mathbf{q})
=
f_{\theta}
\left(
\mathbf{q},
\sum_{i=1}^{N}
\alpha_i(\mathbf{q})y_i
\right),
\label{eq:attn-rem}
\end{equation}
where $f_{\theta}(\cdot)$ denotes a learnable prediction function.

This formulation differs from IDW and Nadaraya--Watson regression in that the query--measurement compatibility is learned from data rather than determined by a predefined distance or kernel function. The resulting attention weights therefore represent task-dependent relevance to the current query, rather than calibrated measurement reliability.

Given supervised training queries $\{\mathbf{q}_m\}_{m=1}^{M}$ with corresponding field values $\{g(\mathbf{q}_m)\}_{m=1}^{M}$, the model parameters can be learned by minimizing:
\begin{equation}
\label{prob:atten-rem-loss}
\mathcal{L}(\theta)
=
\frac{1}{M}
\sum_{m=1}^{M}
\left(
\hat g(\mathbf{q}_m)
-
g(\mathbf{q}_m)
\right)^2.
\end{equation}

\subsubsection{Illustrative Comparison}
We consider a synthetic two-dimensional radio field over $\Omega=[0,1]^2$ and randomly sample a sparse set of noisy measurements from the field. Additional query locations from the same synthetic environment are used to train the learnable attention estimator through \eqref{prob:atten-rem-loss}. For illustration, we compare the resulting estimator with IDW and Gaussian-kernel NW regression, representing fixed distance-based and fixed-kernel aggregation, respectively.

\begin{figure*}[!t]
\centering
\includegraphics[width=0.96\textwidth]{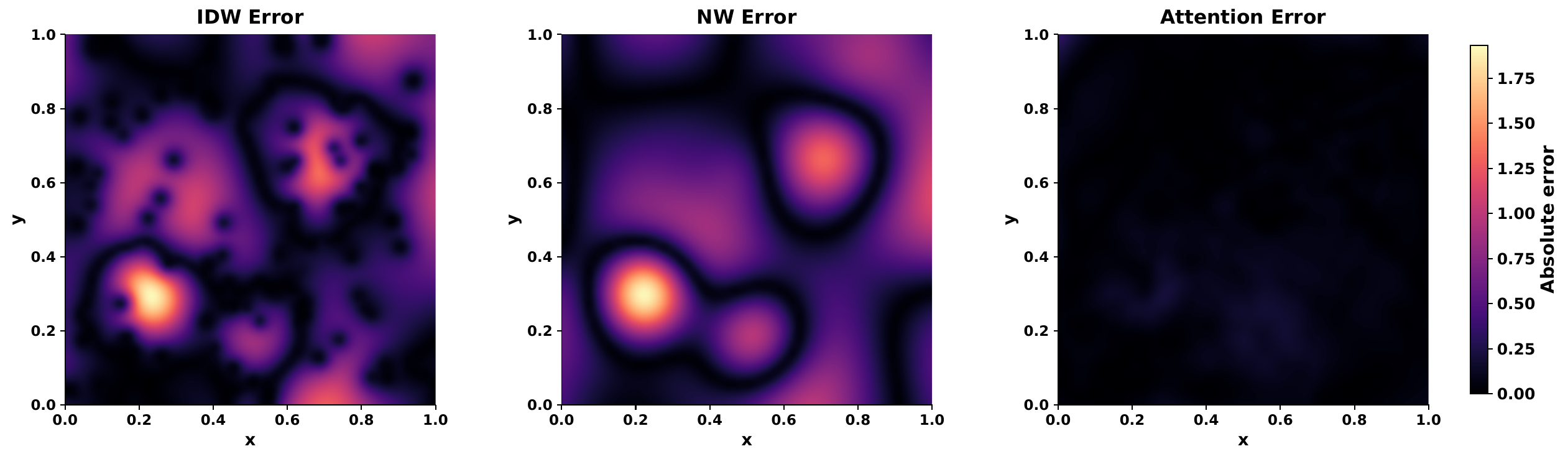}
\caption{Illustrative absolute reconstruction error maps for IDW, NW regression, and a learnable attention-based estimator on a synthetic radio field. The example visualizes qualitative differences among fixed distance weighting, fixed-kernel weighting, and learned query-conditioned aggregation; it is not intended as a general performance benchmark.}
\label{fig:attn_rem_error_maps}
\end{figure*}

\begin{figure}[!t]
\centering
\includegraphics[width=0.4\textwidth]{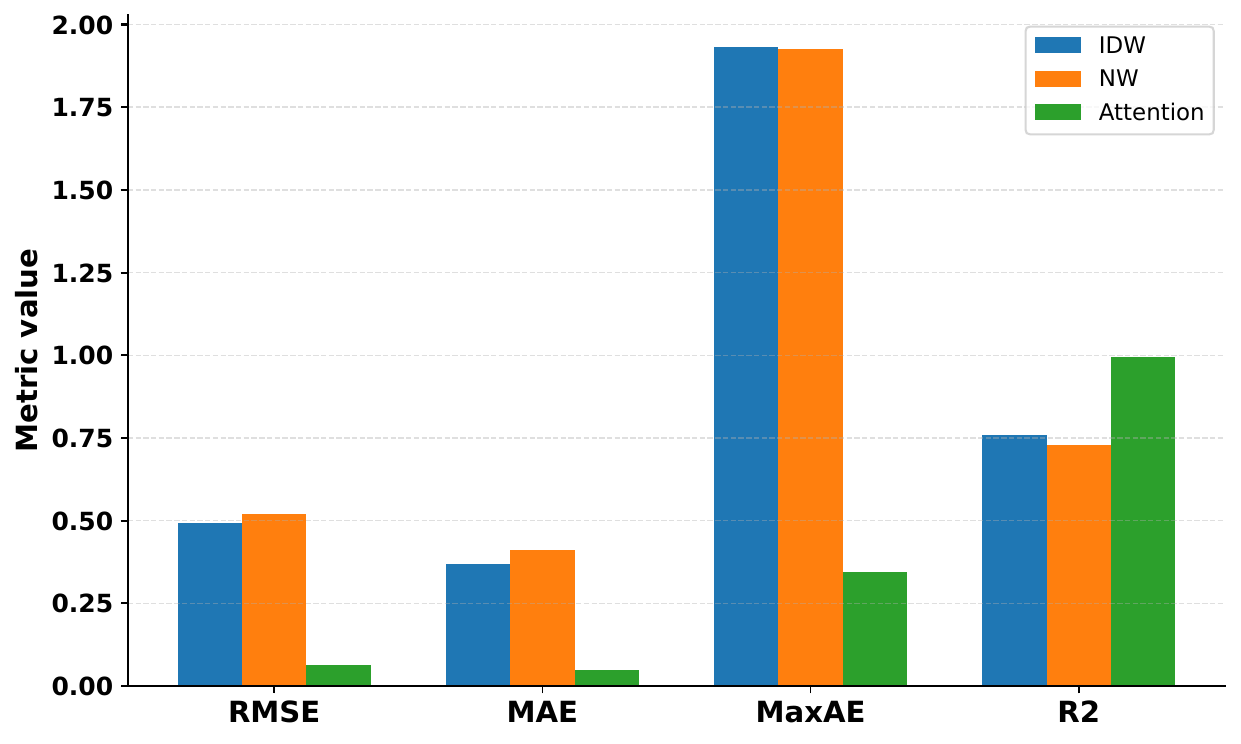}
\caption{Quantitative results for the illustrative synthetic radio-map experiment. The learnable attention estimator attains lower reconstruction errors than IDW and NW for this particular setup while achieving a higher $R^2$. These results illustrate the potential benefit of learned query-dependent weighting rather than establishing general superiority across radio-map environments.}
\label{fig:attn_rem_metric_bar_chart}
\end{figure}

Reconstruction quality is evaluated using root mean square error (RMSE), mean absolute error (MAE), maximum absolute error (MaxAE), and the coefficient of determination ($R^2$). Fig.~\ref{fig:attn_rem_error_maps} visualizes the spatial distribution of reconstruction errors, while Fig.~\ref{fig:attn_rem_metric_bar_chart} summarizes the corresponding quantitative metrics.

For the considered synthetic field, the learnable attention estimator produces smaller and more spatially uniform reconstruction errors than IDW and NW, particularly around regions with stronger spatial variation. The quantitative results in Fig.~\ref{fig:attn_rem_metric_bar_chart} show the same trend in RMSE, MAE, MaxAE, and $R^2$.
Unlike IDW and NW, the attention-based estimator is trained using additional supervised query--field pairs from the same synthetic environment. Its advantage therefore reflects both the flexibility of learned query-dependent weighting and access to training data that the non-parametric baselines do not require. The comparison is intended to illustrate the conceptual difference between fixed and learned aggregation rather than to provide a fully controlled benchmark among reconstruction paradigms.

\subsubsection{Interpretation and Limitations}
The example highlights a progression from fixed to learnable weighting. IDW assigns observation influence according to a predefined distance-decay rule, NW regression replaces this rule with a prescribed similarity kernel, while learnable attention adapts the query--measurement compatibility from data. This additional flexibility can capture spatial relationships that are difficult to encode through a fixed kernel, but it also introduces dependence on training distribution, model capacity, and optimization.

Several limitations should therefore be noted. First, the experiment uses a single synthetic field and matched training and evaluation conditions, so the observed performance may not transfer to different propagation environments, measurement densities, or noise distributions. Second, the attention weights in \eqref{eq:attn-rem} quantify task-dependent relevance and should not be interpreted as calibrated measurement reliability. Third, the comparison does not control for differences in supervision, since the learnable model uses training labels while IDW and NW are training-free estimators. More comprehensive evaluation against learning-based and probabilistic baselines under realistic propagation data would be required before drawing broader conclusions.
Accordingly, this example is a pedagogical illustration of query-conditioned attention rather than as a benchmark result. Its main purpose is to show how spectrum-cartography inference can move from predefined spatial weighting toward learned, context-dependent aggregation.

\section{CONCLUSION}
\label{sec:conclusion}
This survey reviewed learning-based spectrum cartography for low earth orbit (LEO) satellite networks from an integrated inference--decision perspective, covering radio map reconstruction, localization and state estimation, and map-informed resource allocation. Across these tasks, the central challenge is to extract useful spatial radio knowledge from sparse, heterogeneous, and time-varying observations while accounting for propagation context, geometry, and uncertainty. We examined the evolution from classical interpolation, statistical estimation, and structured optimization toward learning-based approaches that exploit environmental context, relational structure, generative modeling, and adaptive fusion. Attention was discussed as an important mechanism within this broader landscape, particularly for learning query- and context-dependent interactions among irregular observations rather than directly estimating measurement reliability. Several common lessons emerge: no single paradigm is universally preferable; classical and model-based methods remain attractive when data are limited or reliable priors are available, while learning-based methods offer greater flexibility under complex conditions; representation and contextual information are often as important as model architecture; and generalization, uncertainty calibration, map freshness, and scalable inference remain key challenges in dynamic LEO environments.
Future work should focus on physics--learning integration, uncertainty calibration, generalization, scalable distributed inference, realistic evaluation, and tighter inference--decision coupling.



\bibliographystyle{IEEEtran}
\bibliography{LEO_Survey}


\end{document}